\documentclass[preprintnumbers, floatfix, preprintnumbers, letterpaper, superscriptaddress,nofootinbib]{revtex4-2}
\pdfoutput=1
\usepackage{graphicx}
\usepackage{microtype}
\usepackage{amsmath}
\usepackage{amssymb}
\usepackage{subfigure}
\usepackage{hyperref}
\usepackage{url}
\usepackage{xcolor}
\usepackage{color}
\usepackage{mathrsfs}
\usepackage{calrsfs}
\usepackage{amsfonts}
\usepackage{latexsym}
\usepackage{ragged2e}
\usepackage{epstopdf}
\usepackage{textcomp}
\usepackage{phaistos}
\usepackage{ulem}

\makeatletter
\renewcommand\@makefnmark{\hbox{\@textsuperscript{\normalfont\color{purple}\@thefnmark}}}
\renewcommand\@makefntext[1]{%
  \parindent 1em\noindent
            \hb@xt@1.8em{%
                \hss\@textsuperscript{\normalfont\@thefnmark}}#1}
\makeatother

\usepackage{caption}
\DeclareCaptionJustification{justified}{\leftskip=0pt \rightskip=0pt \parfillskip=0pt plus 1fil}
\definecolor{vividviolet}{rgb}{0.62, 0.0, 1.0}
\definecolor{amaranth}{rgb}{0.9, 0.17, 0.31}
\definecolor{palatinateblue}{rgb}{0.15, 0.23, 0.89}
\definecolor{brightpink}{rgb}{1.0, 0.0, 0.5}
\definecolor{cornflowerblue}{rgb}{0.39, 0.58, 0.93}
\definecolor{deepcarminepink}{rgb}{0.94, 0.19, 0.22}
\definecolor{radicalred}{rgb}{1.0, 0.21, 0.37}

\hypersetup{ linktoc=all,
    colorlinks, linkcolor={palatinateblue},
    citecolor={brightpink}, urlcolor={amaranth}
}

\graphicspath{{Images/}}

\def\sideremark#1{\ifvmode\leavevmode\fi\vadjust{\vbox to0pt{\vss
 \hbox to 0pt{\hskip\hsize\hskip1em
 \vbox{\hsize1.5cm\tiny\raggedright\pretolerance10000
 \noindent #1\hfill}\hss}\vbox to8pt{\vfil}\vss}}}%
\begin{document}

\title{Gravitating Scalarons with Inverted Higgs Potential}

\author{Xiao Yan \surname{Chew}}
\email{xiao.yan.chew@just.edu.cn}
\affiliation{School of Science, Jiangsu University of Science and Technology, Zhenjiang 212100, China}

\author{Kok-Geng \surname{Lim}}
\email{k.g.lim@soton.ac.uk}
\affiliation{Smart Manufacturing and Systems Research Group, University of Southampton Malaysia, Iskandar Puteri 79100, Malaysia}

\begin{abstract}
Previously, a class of regular and asymptotically flat {gravitating scalar solitons} (scalarons) has been constructed in the Einstein--Klein--Gordon (EKG) theory [JHEP \textbf{07} (2008), 094] by adopting a phantom field with Higgs-like potential where the kinetic term has the wrong sign and the scalaron possesses the negative {Arnowitt--Deser--Misner (ADM)} mass as a consequence. In this paper, we demonstrate that the use of the phantom field can be avoided by inverting the Higgs-like potential in the EKG system when the kinetic term has a proper sign, such that the corresponding gravitating scalaron can possess the positive ADM mass. We systematically study the basic properties of the gravitating scalaron, such as the ADM mass, the energy conditions, the geodesics of test particles, etc. Moreover, we find that it can be smoothly connected to the counterpart hairy black hole solutions from our recent work [Phys. Rev. D, \textbf{109}, 064039 (2024)] in the small horizon limit.
\end{abstract}

\maketitle

\section{Introduction}

{Recently, the detection of gravitational waves {resulting} from the merger of binary compact objects by the LIGO-VIRGO-KAGRA collaboration~\cite{LIGOScientific:2016aoc,LIGOScientific:2017vwq,LIGOScientific:2017ync,LIGOScientific:2019fpa,Barack:2018yly,LIGOScientific:2020ibl} and the imaging of shadows {cast by} two supermassive black holes by the Event Horizon Telescope~\cite{EventHorizonTelescope:2019dse,EventHorizonTelescope:2019ths,EventHorizonTelescope:2019pgp,EventHorizonTelescope:2019ggy,EventHorizonTelescope:2022wkp,EventHorizonTelescope:2022xqj,EventHorizonTelescope:2022urf,EventHorizonTelescope:2022exc} are the major astrophysical events that provide an exciting prospect to search for the existence of the hairy black holes and other compact objects without the event horizon such as boson stars, gravitating scalarons, and wormholes from these astrophysical signatures in the \mbox{future~\cite{Cunha:2019ikd,Sengo:2022jif,Blazquez-Salcedo:2018ipc,Azad:2022qqn,Huang:2023yqd}}. {The ultralight bosonic fields with a mass of $10^{-20}$ eV could possibly explain phenomena such as the halo surrounding a supermassive black hole, leading to the formation of a hairy black hole at the center of the galaxy. Ref.~\cite{Cunha:2019ikd} found that the areal radius of the shadow cast by such hairy black holes is compatible with the observations by the EHT on the supermassive black hole M87, where such quantity could only be weakly constrained due to the current precision of EHT. Furthermore, Ref.~\cite{Sengo:2022jif} found that the effect of gravitational lensing from a rapidly rotating vector-boson star, namely the Proca star and its counterpart hairy black hole, is compatible with the observations about supermassive black holes in M87 and Sgr A$^*$ by EHT. On~the other hand, the~scalar, axial~\cite{Blazquez-Salcedo:2018ipc}, and polar quasinormal modes~\cite{Azad:2022qqn} of the Bronnikov--Ellis (BE) wormhole have been studied. These findings indicate that, when the mass of the wormhole is sufficiently small, it may be feasible to differentiate the wormhole from the Schwarzschild black hole during the ringdown phase of gravitational waves. Similarly, the~study of the optical image of the BE wormhole reveals that the shadow on each asymptotic region may exhibit different behavior. This is due to the presence of a light ring on one side of the asymptotic regions, which could mimic the Schwarzschild black hole, while the absence of a light ring on the other side could provide a unique feature for distinguishing the wormhole~\cite{Huang:2023yqd}.} 
}

In {general relativity (GR)}, a localized {gravitating soliton} supported by a scalar field is known as the~scalaron, a~classic example of which would be the {Fisher--Janis--Newman--Winicour (FJNW)} spacetime, which is static, spherically symmetric, and asymptotically flat but which possesses a naked singularity {at $r=0$ in the radial coordinate $r$} in its spacetime~\cite{Fisher:1948yn,Janis:1968zz,Wyman:1981bd,Virbhadra:1997ie}. It can be obtained by  analytically solving the EKG system with a massless real scalar field $\phi$. {{In addition, the~rotation curve of a galaxy measures the radial velocity of visible matter such as stars, dust, and gas as a function of their distance from the galaxy's center. If~a galaxy only consists of visible matter, then the rotation curve is expected to show that the stars closest to the center of the galaxy would move faster than the stars near the galaxy's outer edge. However, the~observations favor the idea that the inner and outer stars move at roughly the same velocity. Hence, a~potential explanation is that the galactic halo affects the motion of the stars. If~we consider a gravitating scalaron as one of the candidates for the dark matter, some review papers~\cite{Lee:1995af,Magana:2012ph,Lee:2017qve,Lalremruati:2022ern} have pointed out that some research works can fit well the rotation curve when they consider a gravitating scalaron as dark matter. Meanwhile, the ``core--cusp'' problem describes the tension between the observations and numerical simulations of the profile for the central density of the galaxy. The~numerical simulations predict that the central density of a galaxy possesses a cusp-like profile typically described by the Navarro--Frenk--White (NFW) density function, which behaves as $1/r$ at small radii, but whose observations favor the flattened central core. Several research works have resolved this tension by considering that the gravitating scalaron resides at the galactic core, which can provide a good fit for the observational data~\cite{Lee:1995af,Magana:2012ph,Lee:2017qve}. Additionally, to the best of our knowledge, the~Milky Way halo has grown by merging with other galaxies, such as Gaia--Sausage--Enceladus, around 10 billion years ago.}}

However, {it is feasible to remove the naked singularity at $r=0$ such that a regular gravitating scalaron can be obtained where the metric and scalar field with their derivatives are finite at $r=0$. Therefore, in this paper, we employ a scalar potential $V(\phi)$ minimally coupled with Einstein gravity, wherein the~introduction of $V(\phi)$ could cause a scalar field to possess a nonzero effective mass $m_{\text{eff}}=\sqrt{2 \mu}$, since we can assume that $V(\phi)$ contains a quadratic term, i.e.,~$\mu \phi^2$ \footnote{The presence of a quadratic term $\mu \phi^2$ in $V(\phi)$ causes a scalar field to decay as the Yukawa-type potential at infinity, $\phi \sim e^{- \sqrt{2 \mu} r}/r$; thus, this gives rise to a scalar field that possesses an effective mass $m_{\text{eff}}=\sqrt{2 \mu}$.}.} For instance, Ref.~\cite{Dzhunushaliev:2008bq} employed a phantom field with the Higgs-like $V(\phi)$, where its kinetic term possesses a reversed sign. As~a consequence, the~corresponding gravitating scalaron possesses the negative ADM mass. They also considered various configurations of compact objects such as phantom balls, phantom wormholes, etc., in the EKG system with the interaction of two massive real scalar fields~\cite{Dzhunushaliev:2007cs,Dzhunushaliev:2016xdt,Dzhunushaliev:2019mde}.

Nevertheless, in~this paper, we consider inverting the Higgs-like $V(\phi)$ in the EKG system to construct a class of globally regular gravitating scalarons that possess the positive ADM mass, such that the kinetic term of a scalar field can possess a proper sign, and~we demonstrate that the existence of gravitating scalarons depends crucially on the profile of $V(\phi)$, which cannot be strictly positive but has to be negative in some regions. Thus, the usage of the phantom field is not necessary and can be avoided. {This implies that the negativity of $V(\phi)$ in some regions can replace the role of the phantom field to sufficiently violate the energy conditions. We propose this idea because we are inspired by our recent work~\cite{Chew:2023olq}, which has employed the corresponding inverted Higgs-like $V(\phi)$ to evade the no-hair theorem to construct the regular and asymptotically flat hairy black hole that can be bifurcated from the Schwarzschild black hole when the scalar field $\phi_H$ is non-trivial at the horizon \footnote{Previously, Ref.~\cite{Gubser:2005ih} considered such hairy black holes but did not study the properties systematically.}. {The properties of the hairy black holes in our recent work~\cite{Chew:2023olq} are briefly described here}. {As shown in Appendix B (Sec.~\ref{ApB}), {$\phi_H$ can assume any positive real value in principle}; the hairy black hole possesses the positive ADM mass; the~mass and the Hawking temperature increase with an increase in $\phi_H$ { (as shown in Figure~\ref{plot_M_TH})}; its solution and the corresponding derivatives {(as shown in Equations~\eqref{m_ex}--\eqref{p_ex})}, the Ricci scalar {(Equation~\eqref{R_ex})}, and the Kretschmann scalar {(Equation~\eqref{K_ex})} are regular at the horizon; and the violation of the weak energy condition becomes more serious with the increase in $\phi_H$ {(as shown in Figure~\ref{plot_Ttt})}.} 

{On the other hand, to~the best of our knowledge, the~properties of gravitating scalarons in this paper with the inverted Higgs-like $V(\phi)$ have not been reported yet. We have only noticed recently that these gravitating scalarons have been considered in the anti-de~Sitter space, where it is relevant to the context of AdS/CFT~\cite{Astefanesei:2022wmi}. Therefore, we intend to study the properties of gravitating scalarons and then report the results in this~paper.

Furthermore, the~existence of hairy black holes could be connected smoothly by the existence of the gravitating scalarons in the small horizon limit. For~instance, the~hairy black hole can be connected smoothly with the gravitating scalaron in the EKG with $V(\phi)$ containing two asymmetric vacua~\cite{Corichi:2005pa,Chew:2022enh}.  Analogously, we also want to study the possible connection of the hairy black hole in the small horizon limit from our recent work~\cite{Chew:2023olq} along with the gravitating scalaron in this paper.}

Other classes of asymptotically flat gravitating scalarons can also be obtained from the occurrence of the tachyonic instability, which is induced by the anticipation of an additional object with a scalar function $f(\phi)$ in a non-minimally coupled way. For~instance, one can non-minimally couple $f(\phi)$ with a source, which can be in the form of a matter field,such as the~Maxwell field in the Einstein--Maxwell scalar theory~\cite{Herdeiro:2019iwl}, or can be in the form of a geometrical term, such as {the Gauss--Bonnet term within the framework of} the {Einstein-scalar--Gauss--Bonnet (EsGB)} theory~\cite{Kleihaus:2019rbg,Kleihaus:2020qwo}. However, their solutions are not completely regular at $r=0$. For~instance, the~scalar field in the EsGB theory diverges at the origin, while other functions are regular~\cite{Kleihaus:2019rbg,Kleihaus:2020qwo}. Other constructions of gravitating scalarons in GR include the Goldstone model~\cite{Radu:2011uj} and so on~\cite{Kleihaus:2013tba,Karakasis:2023hni,Ning:2023edr,Cadoni:2023wxa}. 

This paper is organized as follows. In~Section~\ref{sec:th}, we briefly introduce our basic theoretical setup, which consists of the profile of $V(\phi)$ in the EKG system and the form of the metric ansatz. Then, we briefly address the existence of gravitating scalarons by performing simple analysis using the {Klein--Gordon (KG)} equation. We also derive a set of coupled differential equations and study the asymptotic behavior of the functions at the origin and infinity. Moreover, we briefly derive the effective potentials for the test particles from the geodesics equation against the background of gravitating scalarons. In~Section~\ref{sec:res}, we present and discuss our numerical findings. Finally, in~Section~\ref{sec:con}, we summarize our~work.

\section{Theoretical~Setting} \label{sec:th}
\unskip

\subsection{Theory and Ans\"atze}

{Here, we begin our construction of the gravitating scalaron in the EKG system by considering a scalar potential $V(\phi)$ of a scalar field $\phi$ {that is minimally coupled with gravity}, }
\begin{equation} \label{EHaction}
 S=  \int d^4 x \sqrt{-g}  \left[  \frac{R}{16 \pi G} - \frac{1}{2} \nabla_\mu \phi \nabla^\mu \phi - V(\phi) \right]  \,,
\end{equation}
{where the scalar potential $V(\phi)$ takes an explicit form as a sum of polynomials in $\phi$} \footnote{{Ref.~\cite{Gao:2021ubl} used Equation~\eqref{polyV} in constructing a hairy black hole but did not systematically study its properties.}},
\begin{equation}
  V(\phi)  = \sum_{i=0}^{\infty} k_i \phi^i \,, \label{polyV}
\end{equation}
{and $k_i$ is an arbitrary constant with real value}. If we consider truncating $V(\phi)$ until $i=4$, then $V(\phi)$ reads as follows:
\begin{equation}
  V(\phi)  =  k_4 \phi^4 + k_3 \phi^3 + k_2 \phi^2 + k_1 \phi \,,
\end{equation}
with $k_0=0$ such that $\phi=0$ at the spatial infinity. $V(\phi)$ could take the asymmetric profile with two asymmetric vacua as shown in Figure~\ref{plot_Vphi}a when $k_3 \neq 0$, {such that $V(\phi)$ has been utilized to obtain solutions for hairy black holes} \cite{Corichi:2005pa,Chew:2022enh} and for the fermionic solitonic star
~\cite{DelGrosso:2023dmv}. {When the coefficients of terms with odd powers become zero}, i.e.,~$k_1=k_3=0$, then $V(\phi)$ can be symmetric with two degenerate minima that look like Higgs-like potentials \footnote{Note that the Higgs-like potential has been considered for constructing a class of traversable wormholes in the Einstein-3-Form theory, where the corresponding kinetic term still can possess a proper sign~\cite{Bouhmadi-Lopez:2021zwt}, and where another class of static and rotating traversable wormholes are supported by the complex phantom field~\cite{Dzhunushaliev:2017syc, Chew:2019lsa}.}, such that $V(\phi)$ has been considered with a phantom field to construct the globally-regular gravitating scalaron with negative ADM mass~\cite{Dzhunushaliev:2008bq}. 

\begin{figure}
\centering
\mbox{
(\textbf{a})
 \includegraphics[trim=50mm 170mm 20mm 20mm,scale=0.55]{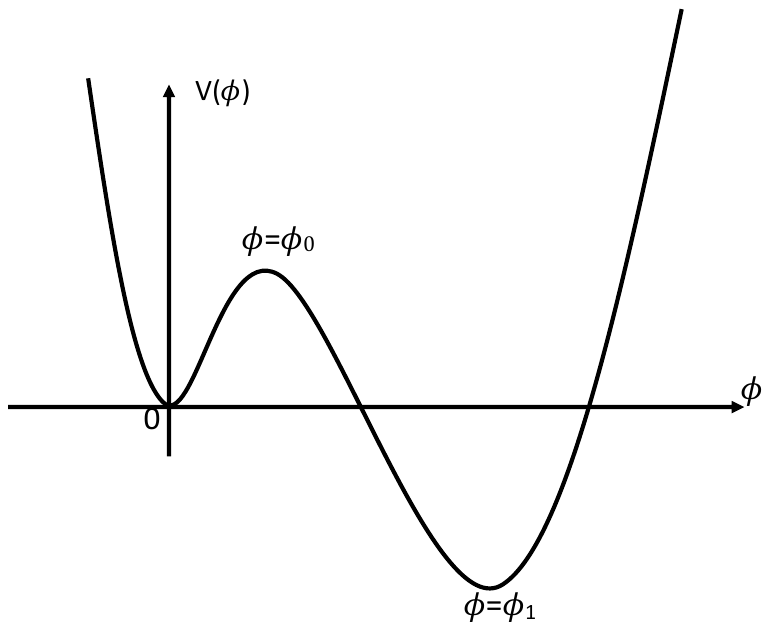}
(\textbf{b})
 \includegraphics[angle =0,scale=0.55]{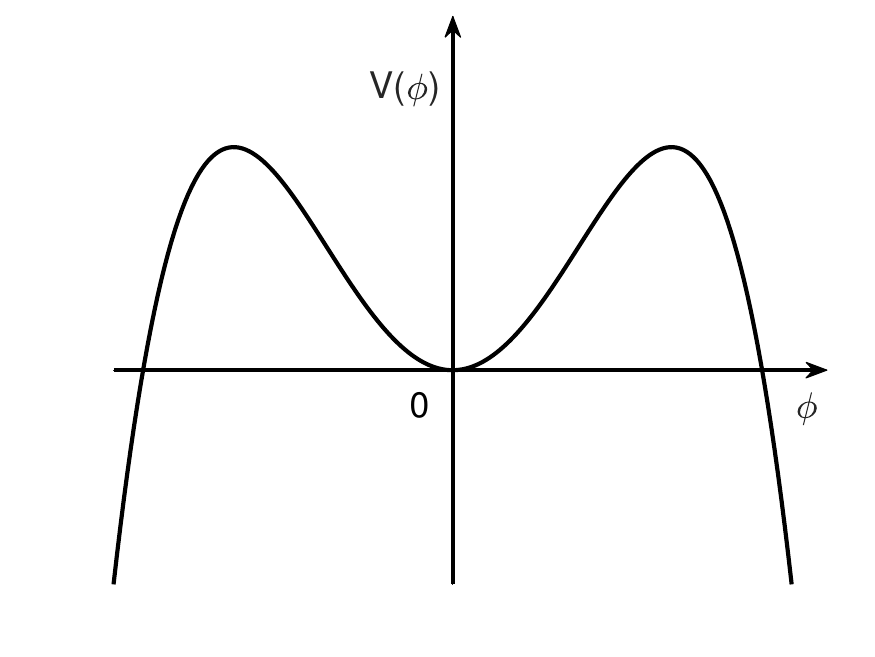}
}
\caption{The illustration of $V(\phi)$: (\textbf{a}) $V(\phi)$ with asymmetric vacua ($\phi=0$ and $\phi=\phi_1$) {has been utilized in constructing hairy black holes} and gravitating scalaron~\cite{Chew:2022enh}. (\textbf{b}) {The inverted Higgs-like $V(\phi)$, which contains two degenerate global maxima at $\phi_{\text{max}}=\pm \sqrt{\mu/(2 \Lambda)}$, has been considered in our recent work for constructing hairy black holes} \cite{Chew:2023olq}.\label{plot_Vphi}}
\end{figure}   

{To achieve} the positive ADM mass of gravitating scalarons in the EKG theory, in this paper, we propose to ``invert'' the Higgs-like $V(\phi)$ such that $V(\phi)$ contains a local minimum at $\phi=0$ and two degenerate global maxima at $\phi_{\text{max}}=\pm \sqrt{\mu/(2\Lambda)}$, {as depicted in Figure~\ref{plot_Vphi}b. The~explicit form of the potential is given by:}
\begin{equation}
 V(\phi) = -\Lambda \phi^4 + \mu \phi^2 \,, \label{ppot}
\end{equation}
{where $\Lambda$ and $\mu$ are constants}. The~above $V(\phi)$ has been considered recently in our previous work~\cite{Chew:2023olq} to construct and then systematically study the basic properties of hairy black hole. Hence, we also want to inspect the possible connection between the hairy black hole and the gravitating scalaron in this~paper. 

{Subsequently, we derive the Einstein equation and the KG equation by varying the action with respect to the metric and scalar fields as follows:}
\begin{align} 
 R_{\mu \nu} - \frac{1}{2} g_{\mu \nu} R &= 8 \pi G \left(  -   \frac{1}{2} g_{\mu \nu} \nabla_\alpha \phi \nabla^\alpha \phi - g_{\mu \nu} V + \nabla_\mu \phi \nabla_\nu \phi    \right) \,, \label{EFE} \\ 
 \nabla_\mu \nabla^\mu \phi  &=  \frac{d V}{d \phi} \,. \label{KGeqn}  
\end{align}
Then, we employ the following ansatz with a spherically symmetric property to describe the gravitating scalaron,
\begin{equation}  \label{line_element}
ds^2 = - N(r) e^{-2 \sigma(r)} dt^2 + \frac{dr^2}{N(r)} + r^2  \left( d \theta^2+\sin^2 \theta d\varphi^2 \right) \,, 
\end{equation}
where $N(r)=1-2 m(r)/r$ with $m(r)$ is the Misner--Sharp mass function~\cite{Misner:1964je}, which yields the ADM mass of gravitating scalarons with $m(\infty)=M$ at infinity. The~scalar field is also assumed to be stationary and spherically symmetric with $\phi \equiv \phi(r)$.

\subsection{Remarks on the Existence of the~Scalaron} \label{ssec:remark}

{Here, the~existence of gravitating scalaron solutions is briefly discussed} by following the discussion from our previous paper~\cite{Chew:2023olq}. In~the introduction part, we have mentioned that, when the scalar field exists in spacetime but the corresponding scalar potential vanishes $(V=0)$, the~solution to the EKG system is the FJNW spacetime~\cite{Fisher:1948yn,Janis:1968zz,Wyman:1981bd,Virbhadra:1997ie} (see Appendix A (Sec.~\ref{ApB})), which is given by
\begin{align}
  ds^2 &= - \left( 1-\frac{b}{r}  \right)^\gamma dt^2 + \left( 1-\frac{b}{r}  \right)^{-\gamma} dr^2 + r^2 \left( 1-\frac{b}{r}  \right)^{1-\gamma} \left( d \theta^2+\sin^2 \theta d\varphi^2 \right)  \,, \label{FJNW} \\
  \phi(r) &= \frac{q}{b} \ln \left(  1-\frac{b}{r}   \right)\,,
\end{align}
where $q$ is the scalar charge, $M$ is the ADM mass, and $\gamma$ is restricted to the range $0 \leq \gamma \leq 1$. These parameters can be related to each other by
\begin{equation}
 \gamma = \frac{2M}{b} \,, \quad b = 2 \sqrt{M^2+q^2} \,.
\end{equation}
Note that, in the limit $\gamma=1$ with $b=2M$, the~Schwarzschild black hole can be recovered from the FJNW metric, where $\phi(r)$ vanishes due to $q=0$. When $0\leq \gamma < 1$, the~spacetime and the scalar field contain a naked singularity in the radial coordinate $r=b$. Hence, it is necessary to properly introduce a non-trivial scalar potential $V(\phi)$ in the EKG system so that the spacetime of gravitating scalarons and the corresponding scalar field can be regular~everywhere. 

Therefore, in~this paper, we employ $V(\phi)$ with an inverted Higgs-like profile to regularize the gravitating scalaron at the origin, so that they are regular everywhere in the spacetime and,~most importantly, so that they possess the positive ADM mass. Recently, we employed the corresponding $V(\phi)$ to construct the hairy black hole, which can be scalarized from the Schwarzschild black hole when the scalar field is non-trivial at the horizon~\cite{Chew:2023olq}. Thus, according to~\cite{Chew:2023olq}, we briefly justify our introduction of $V(\phi)$ in the EKG system, which can indeed guarantee the existence of gravitating scalarons by {simply} referring to the KG equation in performing a few simple analyses without using the Einstein equation~\cite{Herdeiro:2015waa}. {First, we multiply both sides of the KG equation (Equation \eqref{KGeqn}) by $\phi$ and integrate it by parts from the origin to infinity. Subsequently, the~following integral is obtained, given that the boundary term vanishes at both the origin and infinity:}
\begin{equation}
 \int_{0}^{\infty} d^4 x \sqrt{-g} \left[  \nabla_\mu \phi \nabla^\mu \phi + \phi  \frac{d V}{d \phi}   \right] =0 \,.
\end{equation}
Since $\phi$ is spherically symmetric and stationary, then the term $ \nabla_\mu \phi \nabla^\mu \phi = \left( \partial_r \phi \right)^2  \geq 0$. {We further presume that $\phi$ is always positive and nodeless; hence, the condition $\phi \frac{d V}{d \phi} \leq 0$ can be fulfilled if $\frac{d V}{d\phi}= -4 \Lambda \phi^3 + 2 \mu \phi < 0$.} 

Moreover, both sides of Equation \eqref{KGeqn} are multiplied by $\frac{d V}{d \phi}$, and we repeat the above procedure to obtain
\begin{equation}
 \int_{0}^{\infty} d^4 x \sqrt{-g} \left[  \frac{d^2 V}{d \phi^2} \nabla_\mu \phi \nabla^\mu \phi  + \left(   \frac{d V}{d \phi}  \right)^2    \right] = 0   \,.
\end{equation}
The integral above clearly indicates that the condition $\frac{d^2 V}{d \phi^2} < 0 $ must be fulfilled to ensure non-trivial vanishing of the integral. This suggests that the profile of $V(\phi)$ needs to be concave-down and~possibly feature at least one local maximum. Therefore, the~condition $\frac{d^2 V}{d \phi^2} < 0$ can be satisfied if $\frac{d^2 V}{d \phi^2} = -12 \Lambda \phi^2 + 2 \mu  < 0 $. 

On the other hand, the~existence of gravitating scalarons in the EKG system somehow requires the slight violation of energy conditions. {We can examine the {weak energy condition (WEC)}, as it may be violated with $V(\phi) < 0$ in certain regions of $\phi$,}
\begin{equation} \label{WEC}
 \rho = - T^t\,_{t} = \frac{N}{2} \phi'^2 + V  \,.
\end{equation}
Violating the weak energy condition could entail a violation of the strong energy condition, but~the reverse relationship may not necessarily hold true.

\subsection{{Ordinary Differential Equations (ODEs)}} \label{ssec:ode}

{Substituting Equation~\eqref{line_element} into the Einstein equation gives rise to a series of nonlinear ODEs for the functions $m(r)$, $\sigma(r)$, and $\phi(r)$:}
\begin{equation}
m' = 2 \pi G r^2 \left( N \phi'^2 + 2 V \right) \,, \quad \sigma' = - 4 \pi G r \phi'^2 \,,    \quad
\left(  e^{- \sigma} r^2 N \phi' \right)' = e^{- \sigma} r^2  \frac{d V}{d \phi} \,, \label{oode}
\end{equation}
where the prime denotes the derivative of the functions with respect to the radial coordinate $r$. Although~the form of the ODEs looks very simple, it can be very challenging to obtain an analytical solution. {Therefore, we numerically integrate the ODEs from the origin $(r=0)$ to infinity to compute the numerical results, utilizing the professional ODE solver COLSYS~\cite{Ascher:1979iha} and MATLAB package bvp4c} \cite{Shampine:2001}. 

To construct a globally regular gravitating scalaron solution, all functions are required to be finite at $r=0$ by assuming their derivatives vanish at the origin in order to satisfy this regularity condition. The~asymptotic behavior of the functions at the origin can be described by the power series expansion, where a few leading terms are given by:
\begin{align}
 m(r) &= \frac{4 \pi G }{3} V(\phi_c) r^3 + O(r^5) \,, \\
\sigma(r) &= \sigma_c  - \frac{\pi G}{9} \left( \frac{d V(\phi_c)}{d \phi} \right)^2 r^4 + O(r^8)  \,, \\
 \phi(r) &= \phi_c +  \frac{1}{6} \frac{d V(\phi_c)}{d \phi}  r^2 + O(r^4)  \,,
\end{align} 
where $\sigma_c$ and $\phi_c$ are the values of $\sigma(r)$ and $\phi(r)$ at the origin, respectively, {while the condition of asymptotic flatness is imposed at infinity $(r\to \infty)$ by setting the scalar field to vanish. Subsequently, the~leading terms for the corresponding series expansion are expressed as follows:}
\begin{align}
    m(r) &= M  + \tilde{m}_1 \frac{\exp{(- 2 \sqrt{ {2} \mu} r)}}{r} + ...\, ,  \label{in1} \\ 
    \sigma(r) &= \tilde{\sigma}_1 \frac{ \exp{(- 2 \sqrt{ {2} \mu} r)}}{r}  +   ... \, , \label{in2} \\
    \phi(r) &= \tilde{\phi}_{H,1}  \frac{ \exp{(- \sqrt{ {2} \mu} r)} }{r} + ... \, , \label{in3}
\end{align}
where $\tilde{m}_1$, $\tilde{\sigma}_1$, and $\tilde{\phi}_{H,1}$ are constants, and~$M$ is the total mass of the configuration. Moreover, the~effective mass of the scalar field is indicated as $\sqrt{{2}\mu}$. Note that there are a few free parameters, such as $\sigma_c$, $\phi_c$, $M$, $\tilde{m}_1$, $\tilde{\sigma}_1$, $\Lambda$, and $\mu$, in the calculation, where the parameters $\sigma_c$, $\phi_c$, $M$, $\tilde{m}_1$, and $\tilde{\sigma}_1$ are determined exactly when all functions satisfy the boundary conditions on both ends. {Hence, the~following dimensionless parameters are introduced to the ODEs:}
\begin{equation}
 r \rightarrow  \frac{r}{\sqrt{\mu}} \,, \quad m \rightarrow  \frac{m}{\sqrt{\mu}} \,, \quad \phi \rightarrow \frac{\phi}{\sqrt{8 \pi G}} \,, \quad \Lambda \rightarrow 8 \pi G \Lambda \mu \,,
\end{equation}
such that we are only left with $\Lambda$ as the single input parameter in the calculation. For~the sake of numerical calculation, we could also map the one-to-one correspondence of the origin and infinity into $[0,1]$ with the introduction of the compactified coordinate \mbox{$r=x/(1-x)$}. 

In addition, we generate the hairy black hole solutions in this EKG system from our previous work by solving the same set of ODEs in Equation~\eqref{oode} with appropriate boundary conditions (refer to~\cite{Chew:2023olq} for more detail) using COLSYS and MATLAB in order to observe the connection between the hairy black hole and gravitating scalaron in the small horizon~limit. 

\subsection{Geodesics of Test Particles Around the Gravitating~Scalaron} \label{ssec:geo}

{The geodesics of test particles in the vicinity of the gravitating scalaron are studied to understand how the gravitating scalaron could influence the motion of test particles. This investigation could prove useful for future studies aiming to image the shadow of the gravitating scalaron. Nevertheless, we start with the Lagrangian $\mathcal{L}$, defined as follows:}
\begin{equation}
2 \mathcal{L} = \dot{x}^\mu \dot{x}_\mu = \epsilon \,.
\end{equation} 
Here, $\epsilon=0$ corresponds to the massless particle, while $\epsilon=-1$ corresponds to the massive particle. It is worth noting that the dot represents the derivative of a function with respect to an affine parameter.

{As the gravitating scalaron is static and stationary, two conserved quantities arise, namely, the energy $E$ and the angular momentum $L$:}
\begin{equation}
 E = - \frac{\partial {\mathcal{L}}}{\partial \dot{t}} = e^{-2 \sigma} N \dot{t} \,, \quad  L =  \frac{\partial {\mathcal{L}}}{\partial \dot{\varphi}} = r^2 \dot{\varphi} \,. 
\end{equation}
Moreover, we focus on the motion of test particles within the equatorial plane $(\theta=\pi/2)$, and the radial equation for $\dot{r}$ is as follows:
\begin{equation}
 e^{-2 \sigma} \dot{r}^2 =    E^2 -  V_{\text{eff}}(r)  \,, 
\end{equation}
where the effective potential $V_{\text{eff}}(r)$ is given by
\begin{equation}
 V_{\text{eff}}(r) = e^{-2 \sigma} N \left(  \frac{L^2}{r^2} - \epsilon \right) \,.
\end{equation}
We can extract information on the possible appearance of the orbits for the test particles by analyzing $V_{\text{eff}}(r)$.

\section{Results and~Discussions}\label{sec:res}

Here, we present and discuss our numerical results. Figure~\ref{plot_prop1} exhibits the ADM mass $M$ of gravitating scalarons versus the scalar field at the origin $\phi_c$. Analogous to the gravitating scalaron supported by $V(\phi)$ with asymmetric vacua~\cite{Chew:2022enh}, our gravitating scalarons also bifurcate from the Minkowski space by gaining the mass $M$ when $\phi_c$ increases where $M$ is positive. Hence, we demonstrate that the construction of globally-regular gravitating scalarons with a phantom field can be avoided~\cite{Dzhunushaliev:2008bq}, since the profile of $V(\phi)$ is more crucial to determining the existence of gravitating scalarons. It has been systematically discussed in Section~\ref{ssec:remark}. Hence, our gravitating scalaron is less exotic if compared with~\cite{Dzhunushaliev:2008bq}. Meanwhile, the~color bar in Figure~\ref{plot_prop1} depicts the size of horizon $r_H$ of the hairy black hole solutions with a fixed scalar field at the horizon generated from our previous work~\cite{Chew:2023olq}. We observe that the hairy black hole is connected smoothly with the gravitating scalaron in the small horizon limit $(r_H \rightarrow 0)$.
\begin{figure}
 \includegraphics[angle =-90,scale=0.3]{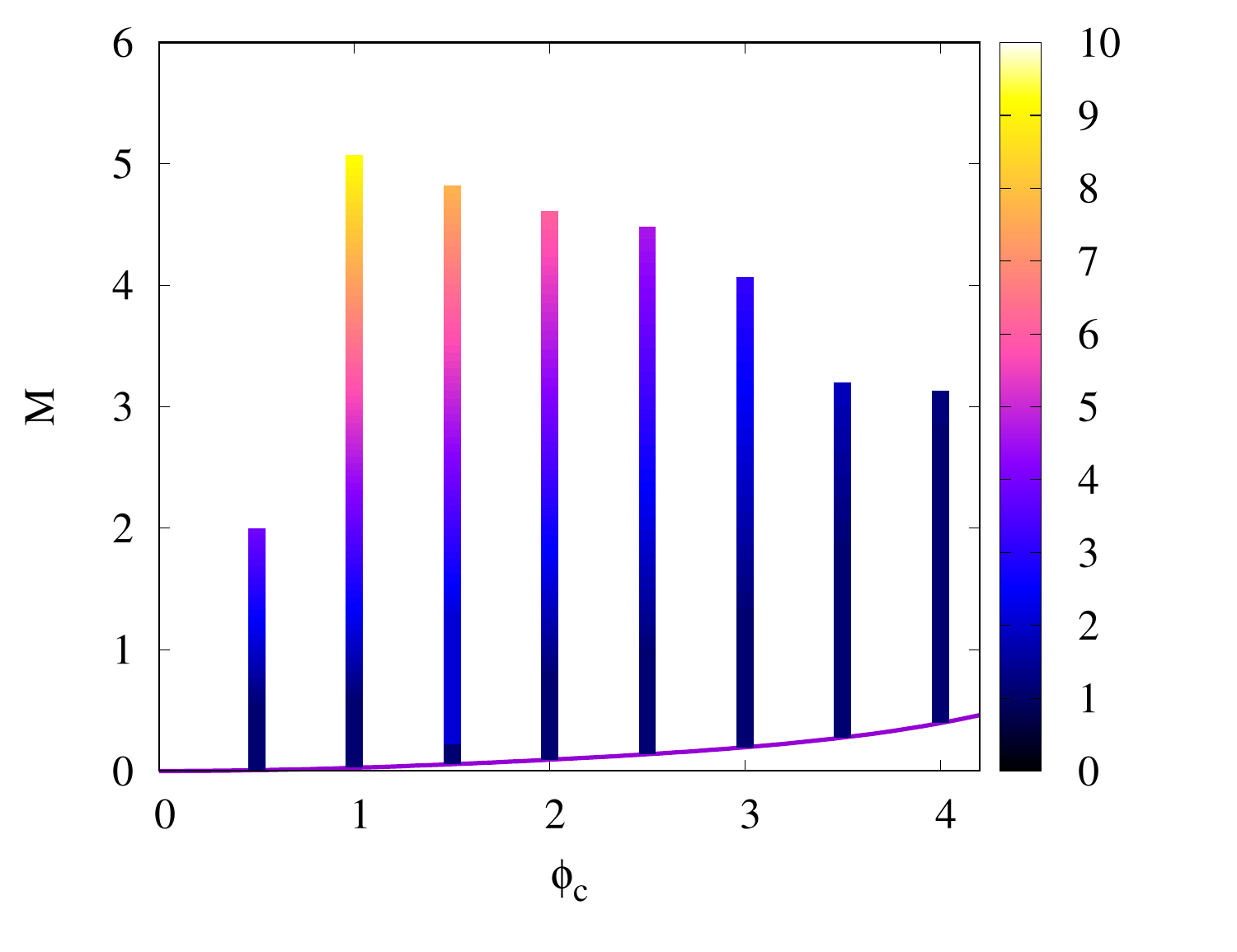}
%
\caption{The basic properties of gravitating scalarons: the ADM mass versus the value of the scalar field at the origin $\phi_c$. Note that the color bar indicates the size of the horizon of the hairy black~holes.}
\label{plot_prop1}
\end{figure}

Analogous to the counterpart hairy black hole~\cite{Chew:2023olq}, Figure~\ref{plot_V}a shows that the coefficient of the quartic self-interaction term {$\Lambda$ can assume any positive real value and that it is inversely proportional to $\phi_c$}. Hence, it is infinitely large when $\phi_c$ approaches zero but decreases monotonically with the increasing of $\phi_c$. {Similarly, Figure~\ref{plot_V}b illustrates that the profile of $V(\phi)$ initially contains two closely spaced degenerate global maxima with very small heights for small $\phi_c$ (large $\Lambda$). However, their separation and height increase as $\phi_c$ increases.} Additionally, $V(0)$ corresponds to the asymptotic flatness of the functions at infinity $(r \rightarrow \infty)$, and the black dot on the yellow curve in Figure~\ref{plot_V}b represents the value of $V(\phi_c)$, which is negative and decreases with the increase in $\phi_c$.

\begin{figure}
\centering 
 \mbox{
 (\textbf{a})
 \includegraphics[angle =-90,scale=0.3]{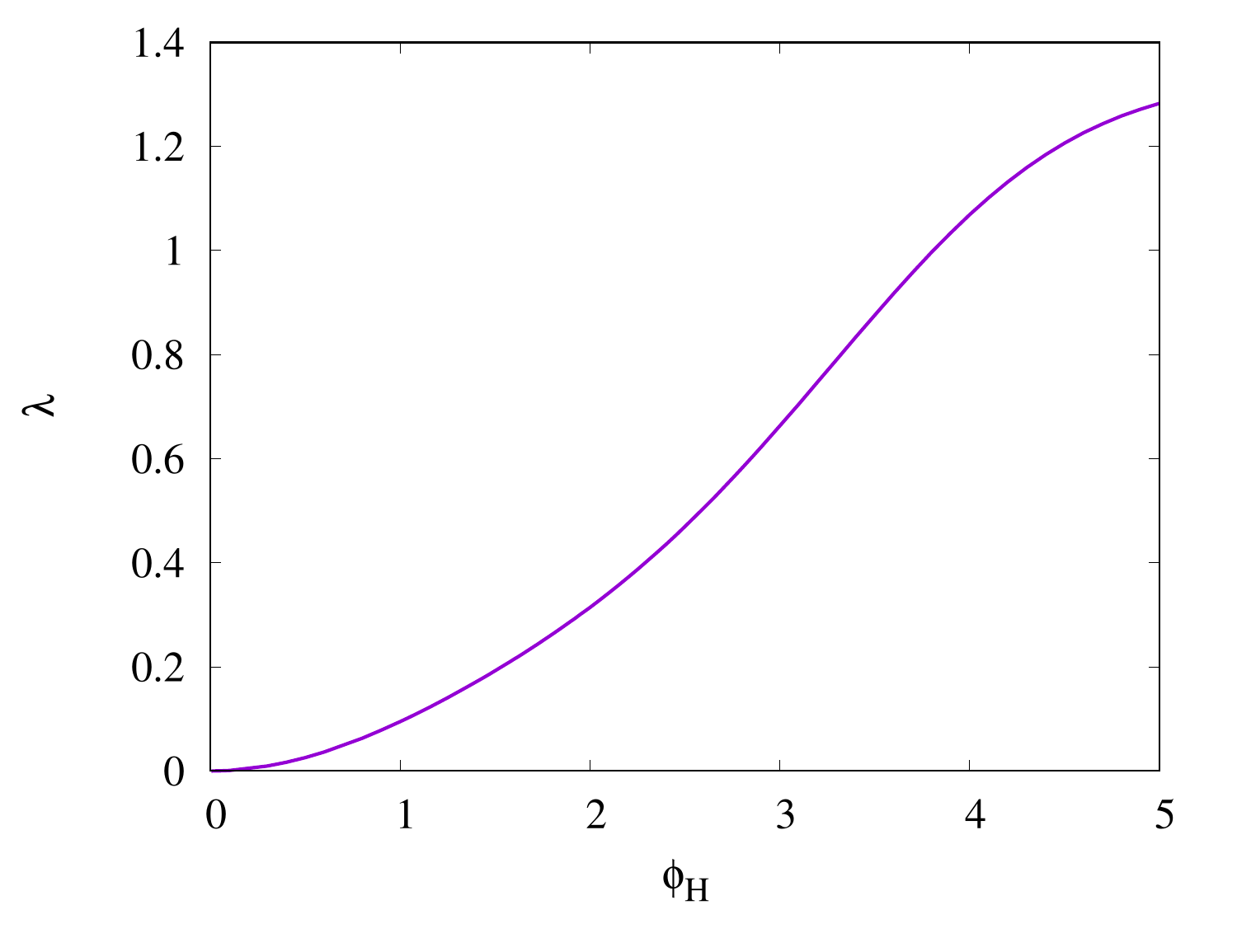}
(\textbf{b})
 \includegraphics[angle =-90,scale=0.3]{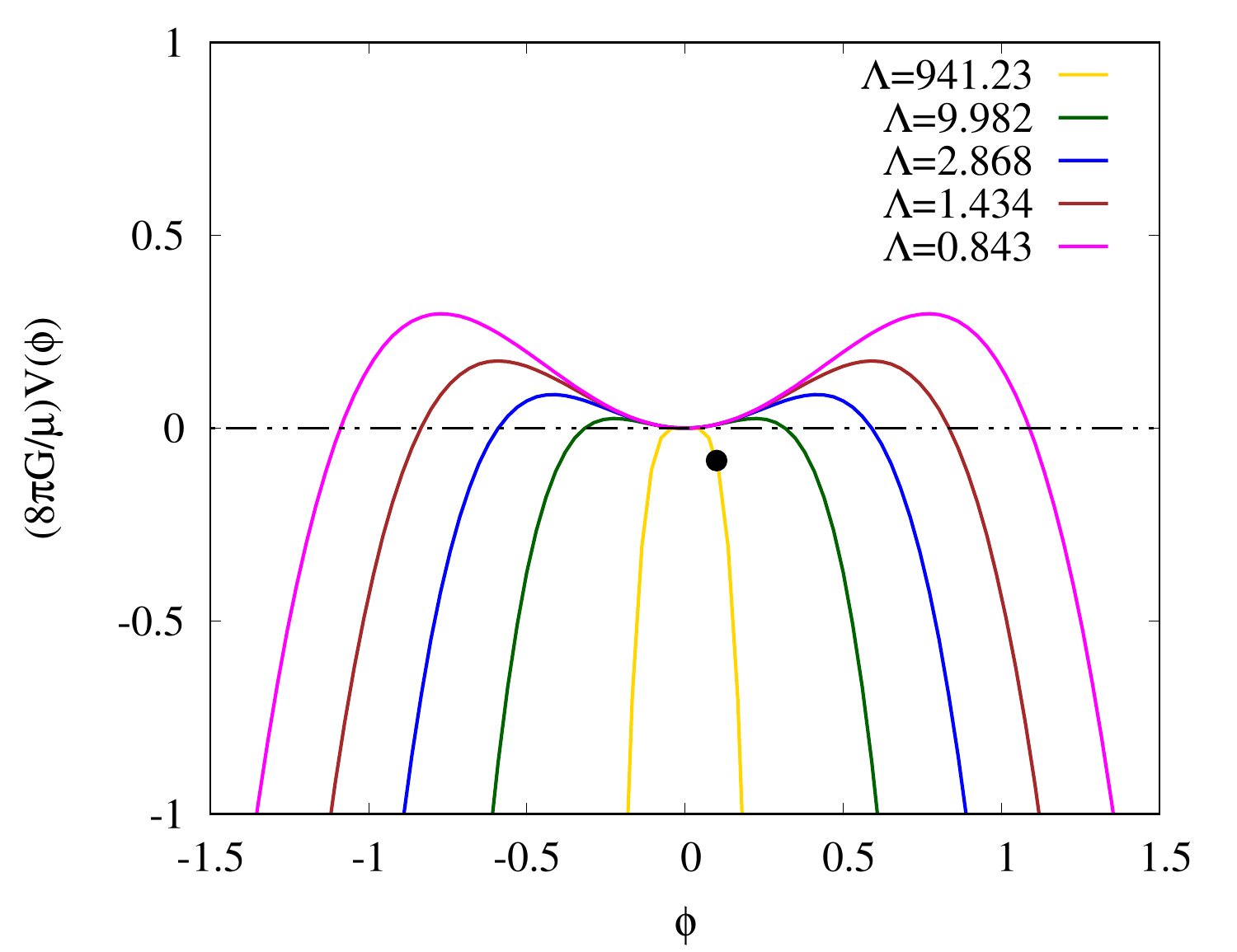}
 }
%
\caption{(\textbf{a}) The value of $\Lambda$ in the logarithmic scale versus $\phi_c$. (\textbf{b}) The profiles of $V(\phi)$ with several values of $\Lambda$. The~black dot on the yellow curve $(\Lambda=941.23)$ represents the value of $V(\phi_c)$.}
\label{plot_V}
\end{figure}

{If we consider a supermassive black hole as a hairy black hole with mass $M_\text{BH}$, immersed in a dark matter halo near a galactic center, then the corresponding density profile $\rho_\text{DM}$ could be described in the form of the power law, $\rho_\text{DM}=\rho_0 (r_0/r)^\gamma$, with the core density $\rho_0$, center radius $r_0$, and power-law index $\gamma$ \cite{Gondolo:1999ef,Nishikawa:2017chy,Kavanagh:2020cfn}. This gives rise to the existence of the dark matter spike with a radius $R_{\text{sp}}(\gamma,M_\text{BH})$,
\begin{equation}
R_{\text{sp}}(\gamma,M_\text{BH})=\alpha_\gamma r_0 \left( \frac{M_\text{BH}}{\rho_0 r^3_0} \right)^{\frac{1}{3-\gamma}} \,,
\end{equation}
where the normalization constant $\alpha_\gamma$ can be calculated numerically for each $\gamma$. The~dark matter spike is formed as a consequence of an adiabatic growth of the hairy black hole, which increases the central density of the dark matter halo. The~distribution of dark matter in the vicinity of the spike region is described by
\begin{equation}
 \rho^\text{sp}_\text{DM}(r) = \rho_\text{sp} \left(  \frac{R_\text{sp}}{r}  \right)^{\gamma_\text{sp}} \,,
\end{equation} 
where $\gamma_\text{sp}=\left( 9-2\gamma \right)/\left( 4-\gamma \right)$, and~$R_\text{sp}=2 M_\text{BH} \approx 2.95 \left( M_\odot/M_\text{BH}\right)$ is the Schwarzschild radius of a hairy black hole. Note that this dark matter density profile differs from the NFW density profile, which is motivated by the numerical simulations of the collisionless dark matter particles in the galactic halos for~$\gamma=0 ,1$ and $M_\text{BH}= 10^5$ or $10^6 M_\odot$.

Although dark matter has not been detected yet, several theoretical studies on the observational signatures for dark matter have been carried out~\cite{Cunha:2019ikd,Sengo:2022jif,Nishikawa:2017chy,Kavanagh:2020cfn}. As~we have mentioned in the introduction part, if~a galaxy could be described by a galactic halo that harbors a supermassive black hole at its galactic center, and~if the corresponding supermassive black hole could form a hairy black hole that is immersed in the ultralight bosonic field with the mass $10^{-22}$ eV, then it can cast a shadow when a hot gas from a disk is accreted into it, where the optical image could be detected by the EHT. Similarly, the~spinning Proca star and its counterpart hairy black hole could be detected from their effect of gravitational lensing~\cite{Sengo:2022jif}. Additionally, Ref.~\cite{Nishikawa:2017chy} studied the emission of gravitational waves from the dark matter spike that was formed by a primordial black hole, and they found that the direction of the emission  originated from the galactic center and has little correlation with the position in the galaxy. Their results could account for the characterization of primordial black holes as dark matter within a specific mass range, as~detected by the LVK collaboration. Ref.~\cite{Kavanagh:2020cfn} found that, when the density of the dark matter spike is not static but evolves with time, the~orbits of the inspirals for a stellar-mass compact object around a black hole immersed in dark matter could be reduced during~the emission of a gravitational wave, which might be detected by LISA in the \mbox{near future. }
}

We show the profiles of gravitating scalaron solutions in the compactified coordinate $x$ in Figure~\ref{plot_sols}. Overall, they behave quite similarly to the counterpart hairy black hole~\cite{Chew:2023olq}. Figure~\ref{plot_sols}a shows that the profile of mass function $m(x)$ behaves similarly to the profile of $m(x)$ of the counterpart hairy black hole~\cite{Chew:2023olq}, where it possesses an almost constant function that is extended from the origin to an intermediate point in the spacetime and then develops a sharp boundary near that intermediate point so that it is changed to another set of almost constant function indicated as the ADM mass at infinity. Note that the almost constant function at infinity corresponds to the local minimum of $V(\phi)$, which is $V(0)=0$, while another almost constant function at the origin corresponds to $V(\phi_c)$.

Meanwhile, Figure~\ref{plot_sols}b,c depict the profiles of functions $\sigma(x)$ and $\phi(x)$ in the compactified coordinate $x$, where they are also regular at the origin but decrease monotonically to zero at infinity. On~the other hand, we find that the profiles of gravitating scalaron solutions behave quite similar to the hairy black holes and gravitating scalarons, supported by $V(\phi)$ with asymmetric vacua~\cite{Chew:2022enh}. In~addition, our solutions are completely regular everywhere, but other classes of gravitating scalarons are not; for~instance, only the scalar field of ultra-compact gravitating scalarons diverges at the origin in EsGB theory~\cite{Kleihaus:2019rbg,Kleihaus:2020qwo}, but other functions and effective stress-energy tensors are regular~everywhere. 

\begin{figure}

\centering 
\mbox{
 (\textbf{a})
 \includegraphics[angle =-90,scale=0.3]{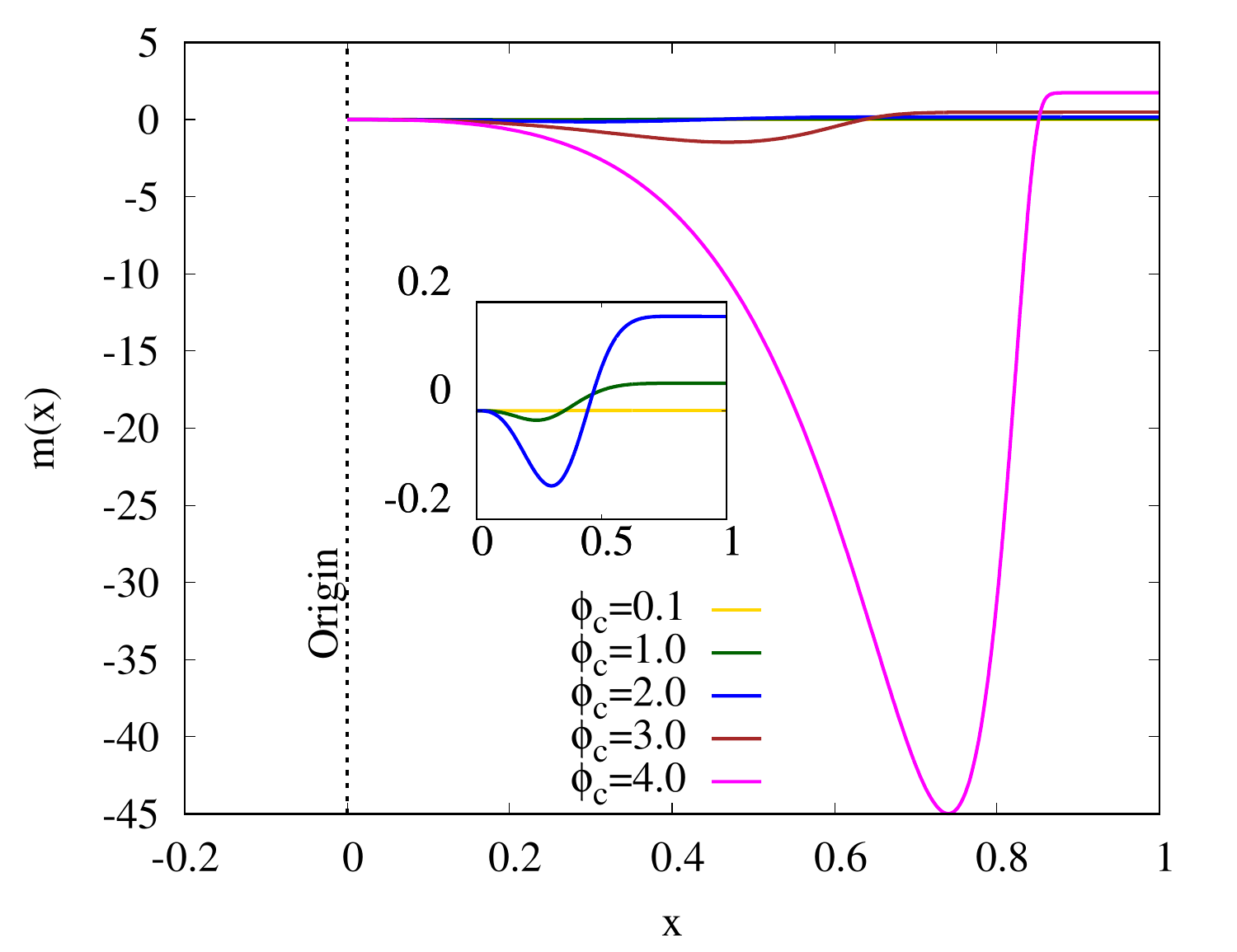}
(\textbf{b})
 \includegraphics[angle =-90,scale=0.3]{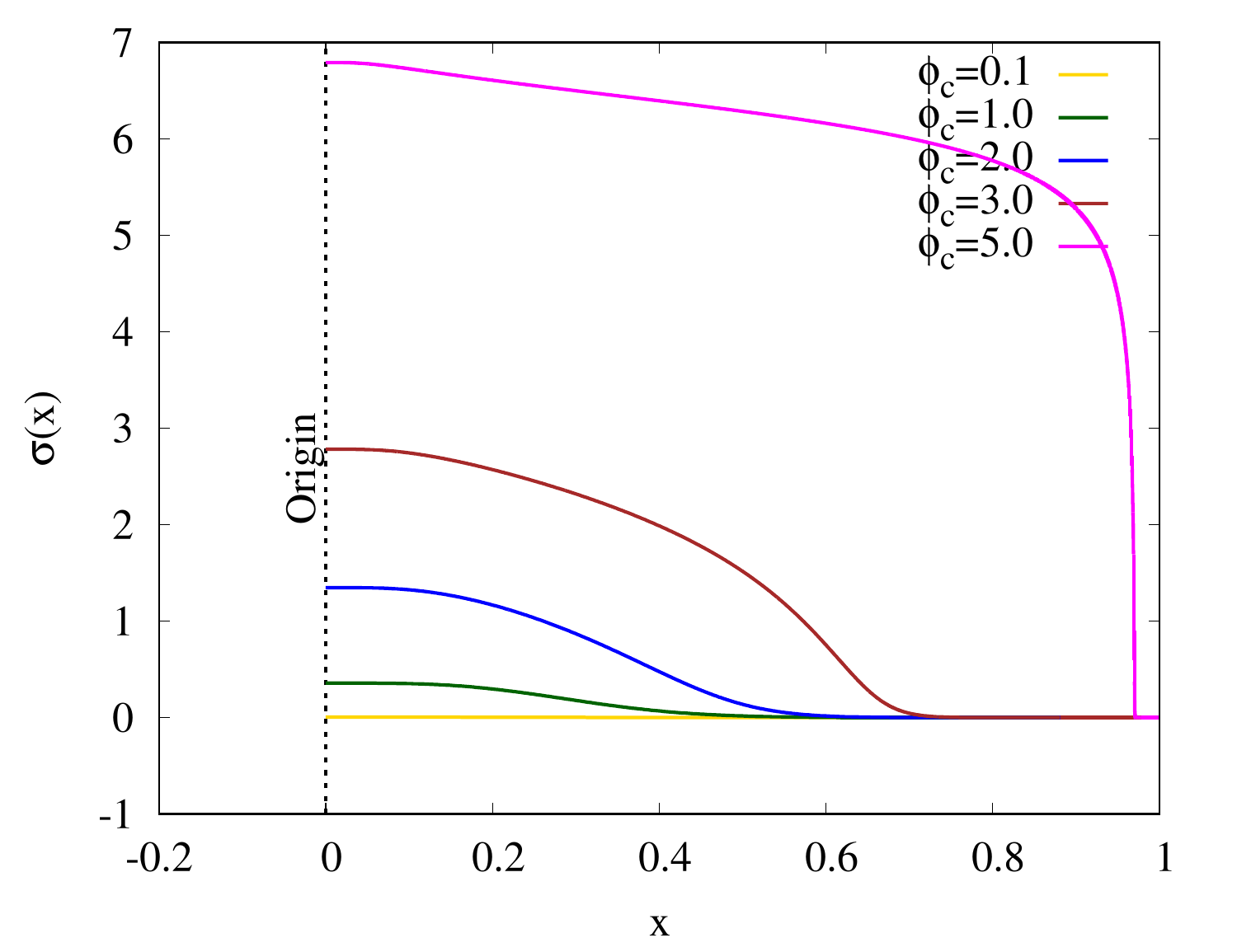}
 }
\mbox{
(\textbf{c})
\includegraphics[angle =-90,scale=0.3]{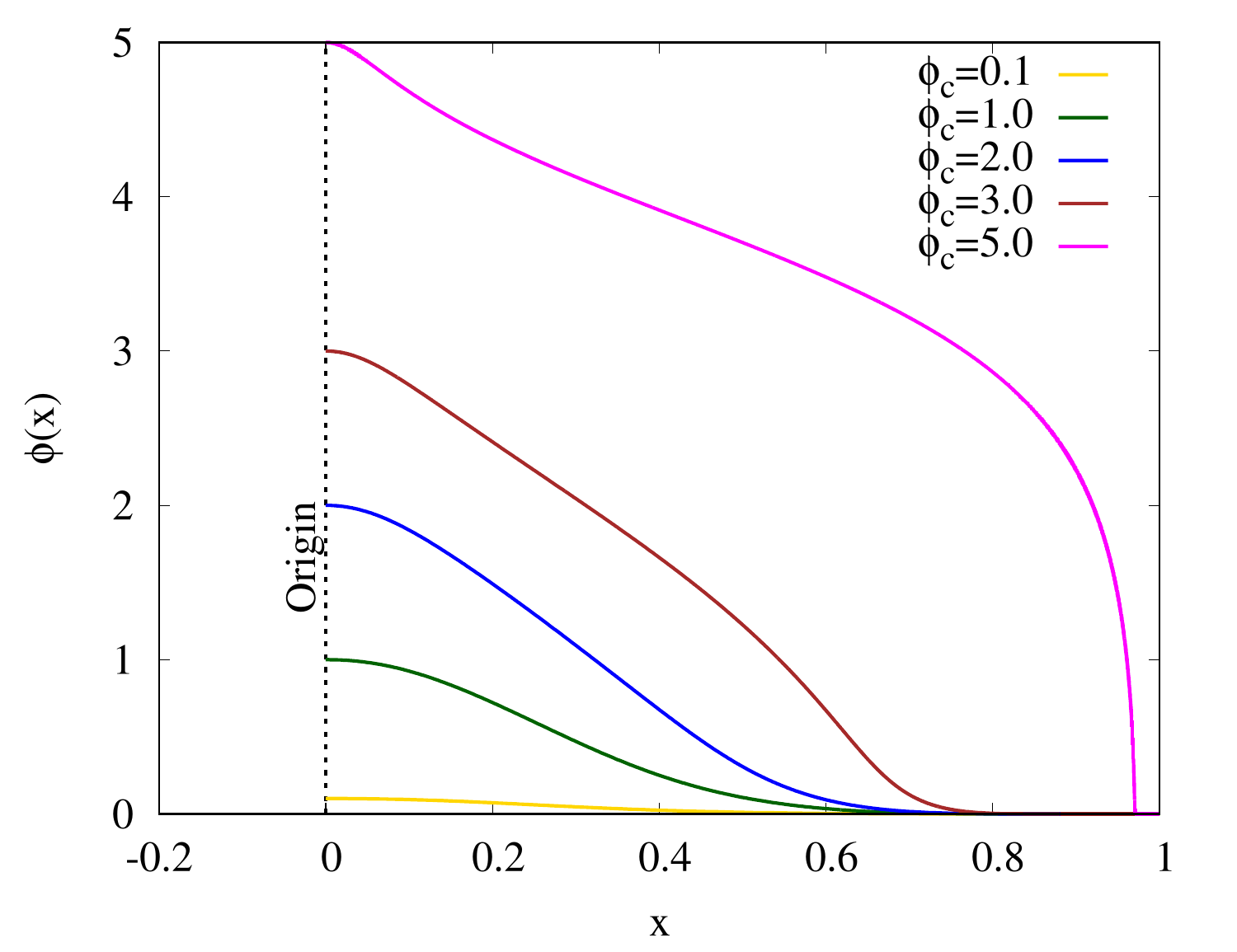}
}
%
\caption{The profiles of gravitating scalaron solutions with several values of $\phi_c$ in the compactified coordinate $x$: (\textbf{a}) $m(x)$; (\textbf{b}) $\sigma(x)$; and (\textbf{c}) $\phi(x)$.}
\label{plot_sols}
\end{figure}

Figure~\ref{plot_energy_cond}a shows the WEC of gravitating scalarons, which is described by $\rho=-T^t\,_t$ in the compactified coordinate $x$. The~WEC is violated, particularly at the origin, and~is severely violated with {the increase in $\phi_c$. Conversely,} another energy condition $T^r\,_r$, depicted by Figure~\ref{plot_energy_cond}b, is being satisfied with the increase in $\phi_c$. Hence, this demonstrates that the profile of Equation~\eqref{ppot} is sufficient to violate the WEC for the existence of regular gravitating scalarons; therefore, the usage of the phantom field for the construction of gravitating scalarons is no longer necessary~\cite{Dzhunushaliev:2008bq}.

\begin{figure}
\centering 
 \mbox{
 (\textbf{a})
 \includegraphics[angle =-90,scale=0.3]{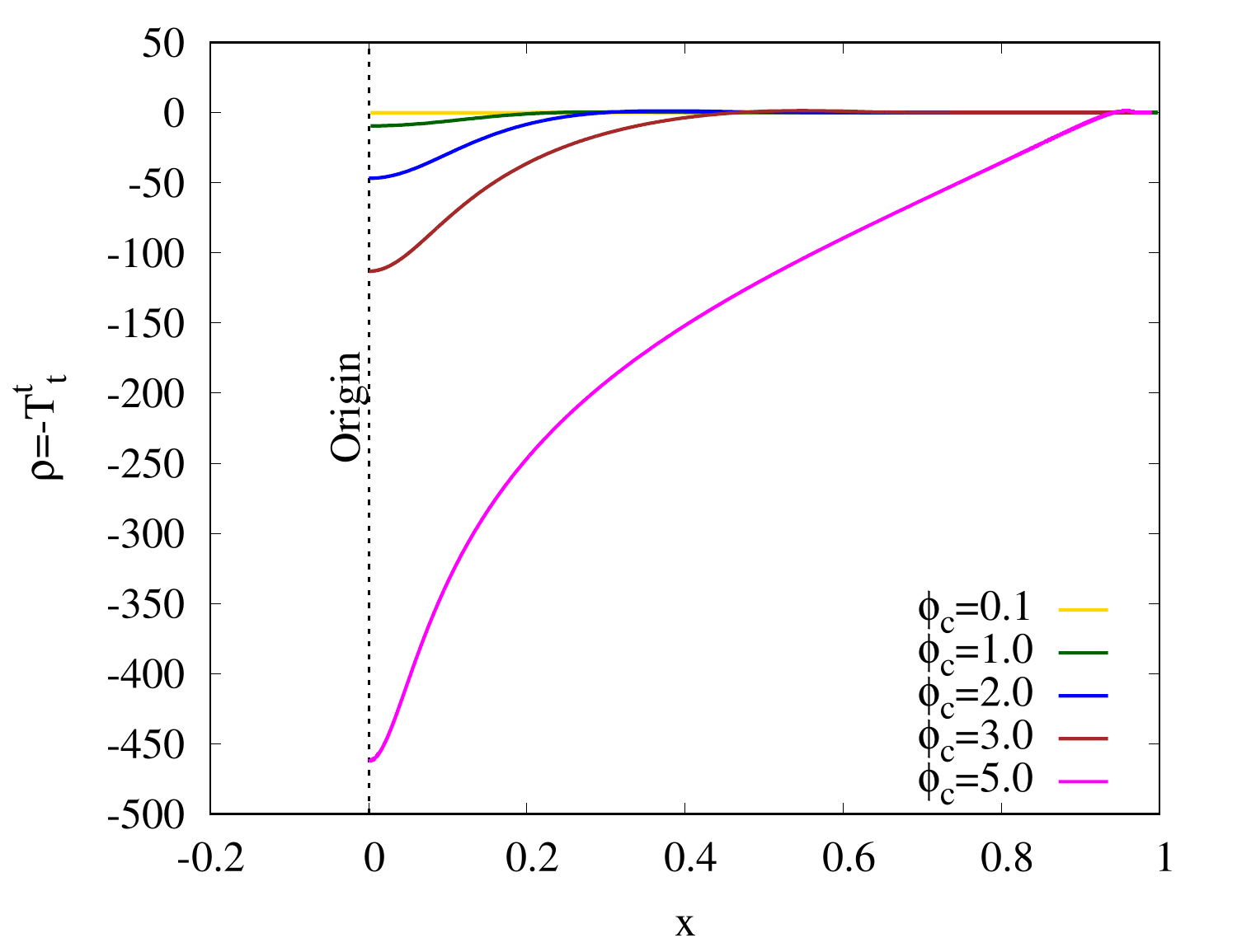}
(\textbf{b})
 \includegraphics[angle =-90,scale=0.3]{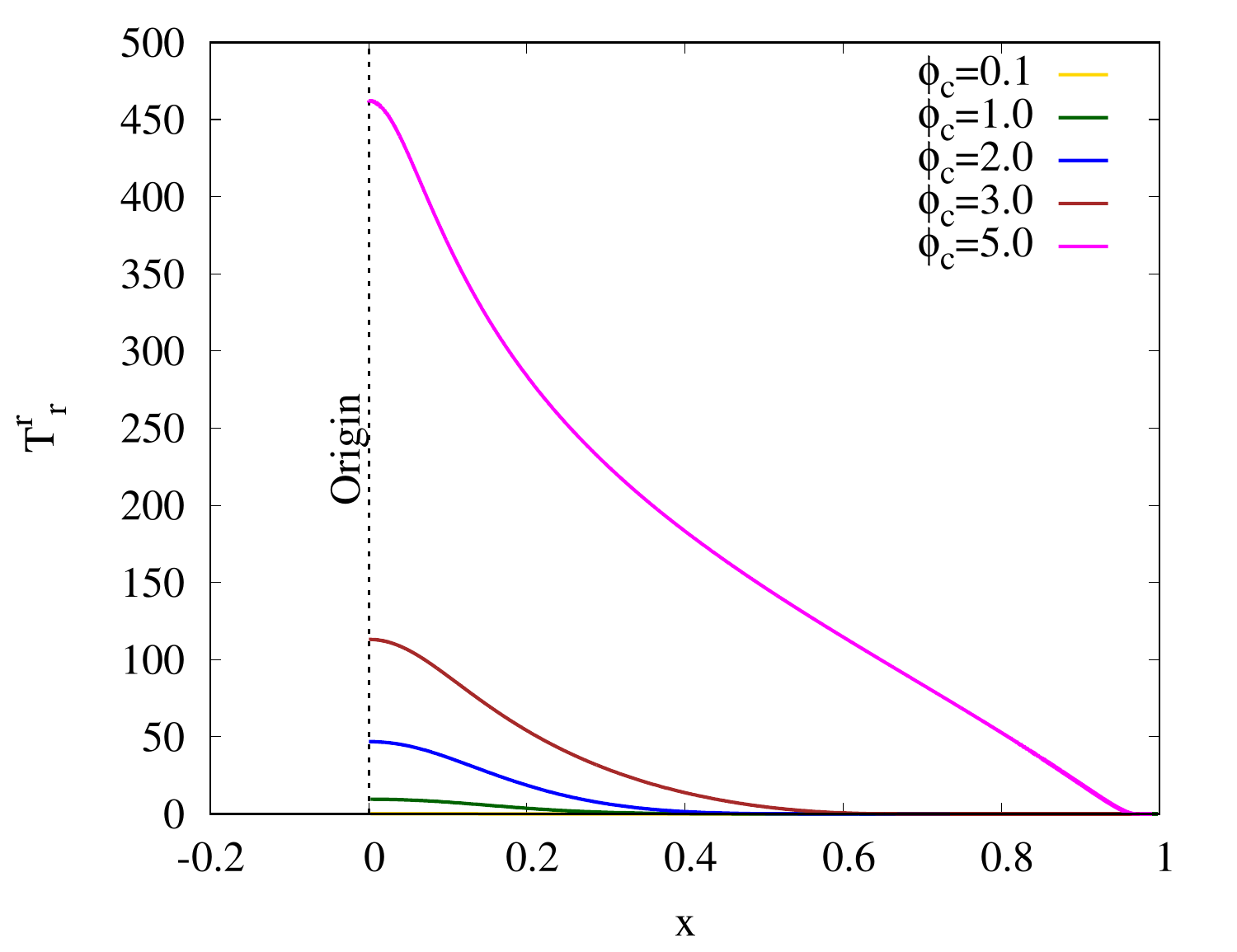}
 }
%
\caption{The scaled energy conditions for the gravitating scalarons with various values of $\phi_c$, depicted as (\textbf{a}) $\frac{8 \pi G}{\mu}\rho$ and (\textbf{b}) $\frac{8 \pi G}{\mu}T^r\,_r$}.
\label{plot_energy_cond}
\end{figure}

We could also gain some insights into the possible appearance of the orbits for a massive test particle in the spacetime of gravitating scalarons by analyzing its scaled effective potential $V_{\text{eff}}(x)/\mu$ in the compactified coordinate $x$ in Figure~\ref{plot_geo}a and \ref{plot_geo}b for $\phi_c=0.5, 3.0$, respectively. When $L=0$, {the massive test particle (black curve) is capable of moving} radially from infinity to reach $x=0$ with $E^2 \geq V_{\text{eff}}(0)/\mu$ or just stays at $x=0$ if it is already at $x=0$. When $L \neq 0$, depending on $E$, it will move either in the bound orbit, which is confined by a minimal radius and a maximal finite radius, or in an escape orbit, which can be extended from a minimal radius to infinity~\cite{Diemer:2013zms}. However, it cannot reach $x=0$ regardless of the values of $E$ and $L$, since $V_{\text{eff}}(x)/\mu$ diverges on $x=0$. Note that it can move in the {innermost stable circular orbit (ISCO)} (magenta curve, $L^2=0.5$) as shown in Figure~\ref{plot_geo}b, where the inner local minimum is indicated as the location of the ISCO. {Since no constraint has been applied to our model from the observations, if~we want to estimate the size of the ISCO, then we adopt a non-rigorous method, which assumes our scalar field will behave similarly to the complex scalar field for the boson star. In~particular, we pick the constraint on the complex scalar field from the boson star with a quartic self-interaction term, since we have it in our theory. Hence, if~we assume the scalar field in our model is a dilaton with mass $10^{-10}$ eV, then the radius of ISCO could be estimated to be on the scale of $10^3$ km. If~our scalar field is the Higgs particle with mass 125 Gev, then the radius of ISCO can be estimated on the scale of $10^{-18}$ m~\cite{Diemer:2013zms, Schunck:2003kk}.}  

\begin{figure}

\centering 
\mbox{
(\textbf{a})
\includegraphics[angle =-90,scale=0.3]{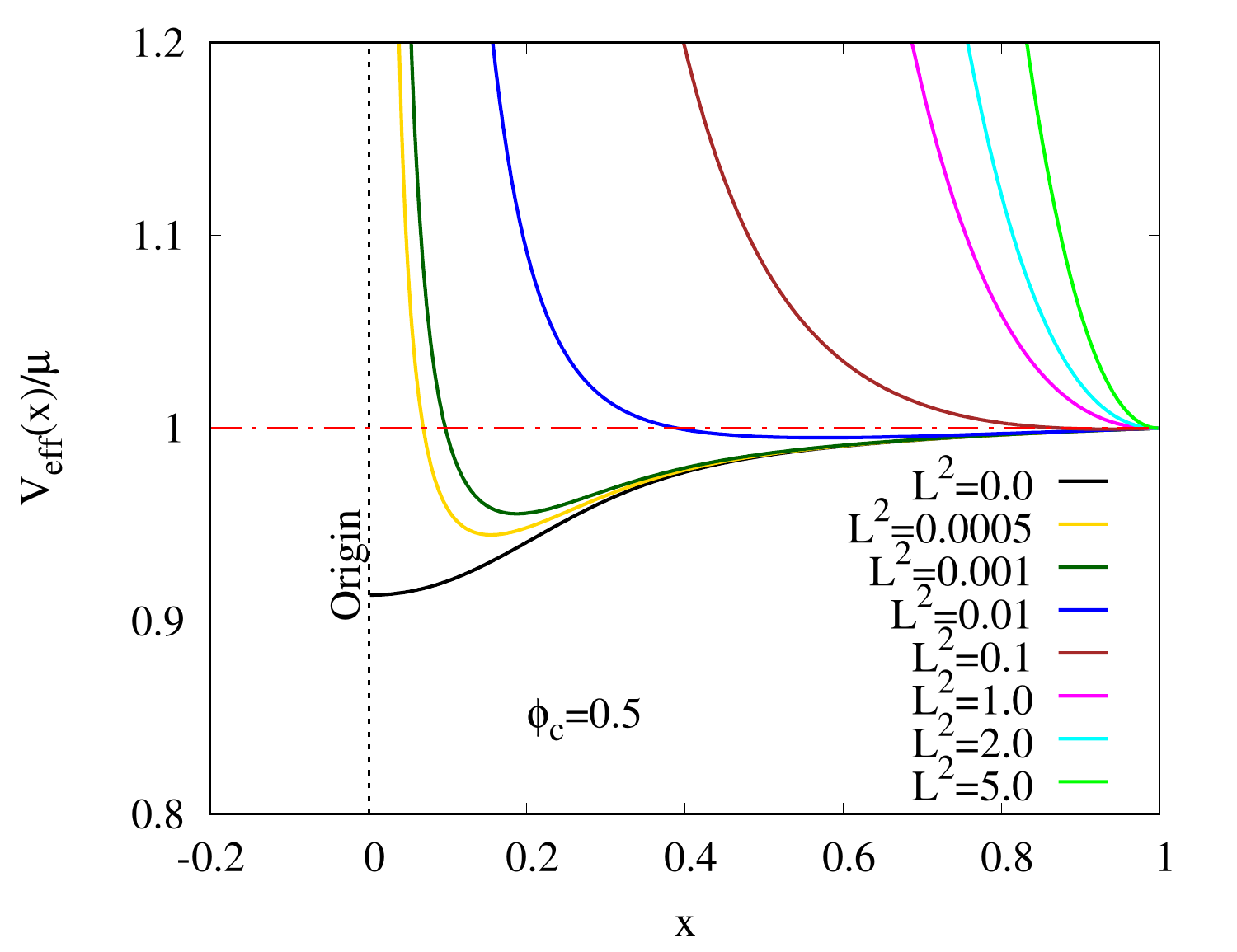}
(\textbf{b})
\includegraphics[angle =-90,scale=0.3]{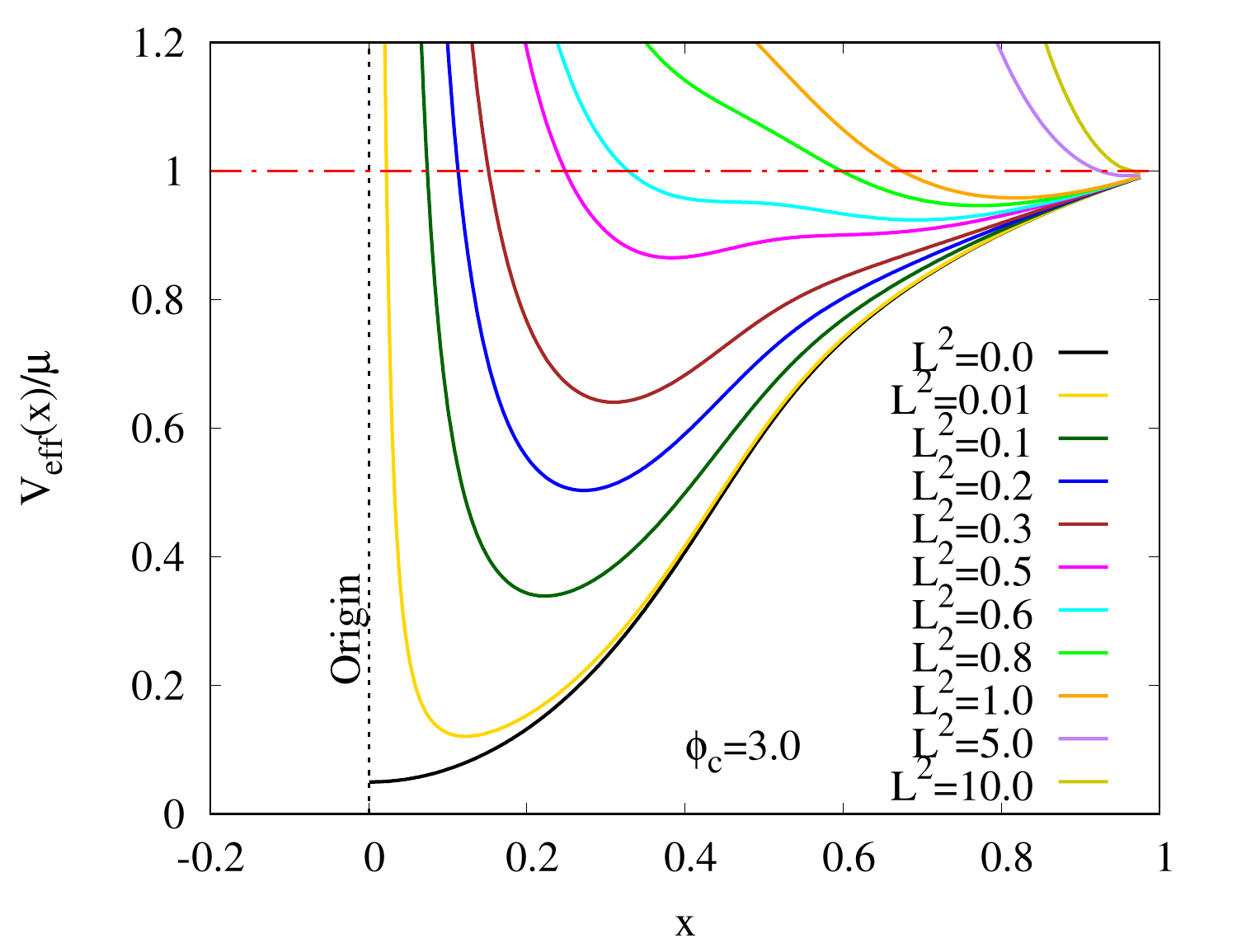}
}
%
\caption{The scaled effective potential $V_{\text{eff}}(x)/\mu$ of a massive particle in the compactified coordinate $x$ with several values of $L^2$ for (\textbf{a}) $\phi_c=0.5$ and (\textbf{b}) $\phi_c=3.0$.}
\label{plot_geo}
\end{figure}

{The scaled effective potential of the light ring $V_{\text{eff}}(x)/(\mu L^2)$ in the compactified coordinate $x$ is depicted in Figure~\ref{plot_geo0}, where it diverges at $x=0$ and monotonically decreases. Thus, there is no light ring in the spacetime.} Overall, we find that the types of orbits appearing in the gravitating scalaron are quite similar to the case of the boson star, which is supported by the {superpotential and supersymmetry (SUSY)} potential~\cite{Diemer:2013zms}.

\begin{figure}

 \includegraphics[angle =-90,scale=0.3]{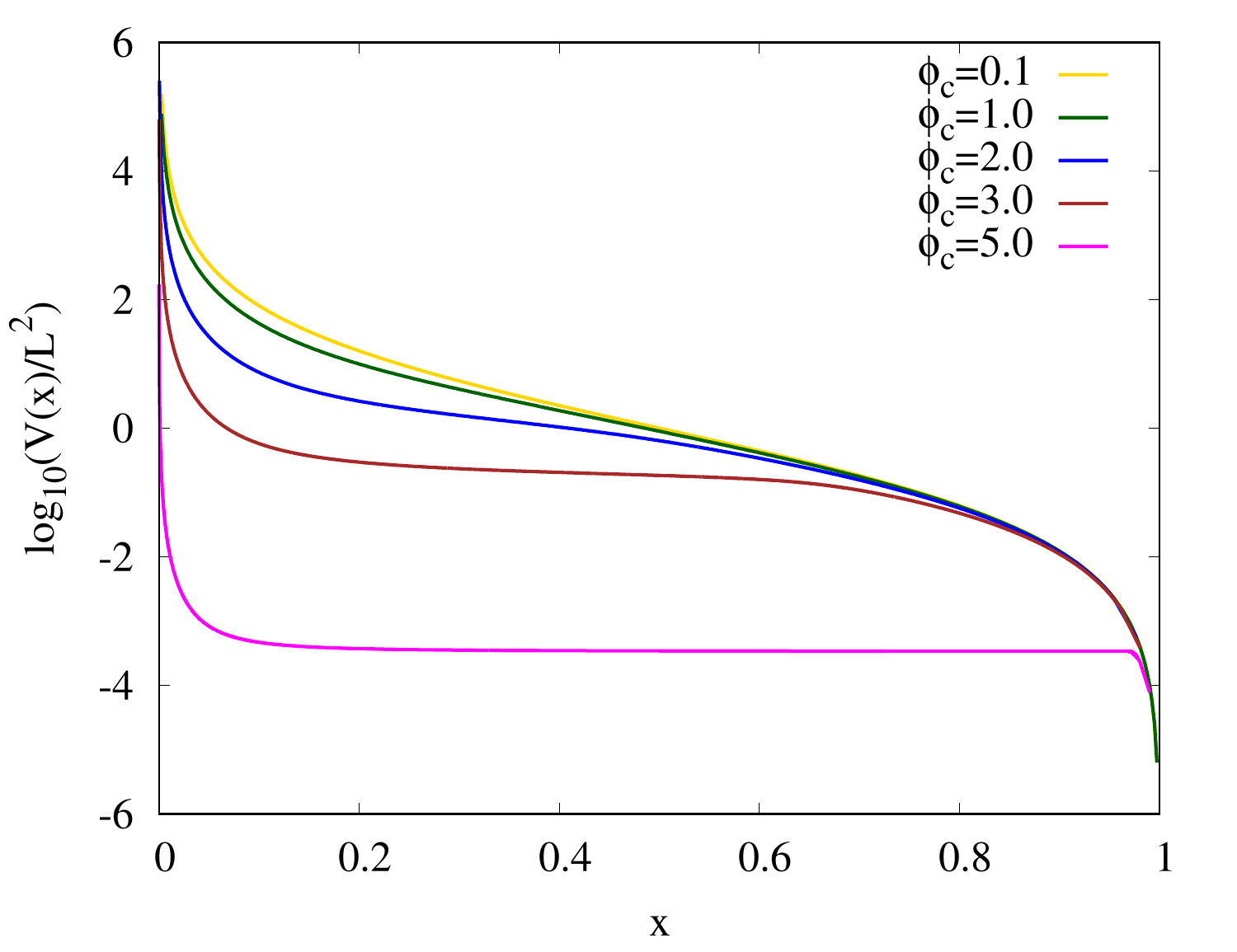}

\caption{The scaled effective potential for the light ring $V_{\text{eff}}(x)/(\mu L^2)$ in the compactified coordinate $x$ with several values of $\phi_c$.}
\label{plot_geo0}
\end{figure}

Finally, we briefly address the issue regarding the linear stability of the gravitating scalaron. Ref.~\cite{Chew:2022enh} reveals that the configurations of hairy black holes and gravitating scalarons supported by $V(\phi)$ with asymmetric vacua are unstable against a small linear perturbation in the background. Since the hairy black hole supported by the inverted Higgs-like potential is unstable against the linear perturbation~\cite{Chew:2023olq}, {and since the hairy black hole can be smoothly connected with our gravitating scalaron in the small horizon limit, we therefore conjecture that our gravitating scalaron could inherit the unstable modes from the hairy black hole in the small horizon limit and that our gravitating scalaron is very likely to be unstable against the linear perturbation.}

\section{Conclusions}\label{sec:con}

A few decades ago, a class of globally regular and asymptotically flat gravitating scalarons was considered in the EKG system, where the Higgs-like potential with the phantom field minimally couples with Einstein gravity. As~a consequence, the~corresponding gravitating scalaron possesses the negative ADM mass~\cite{Dzhunushaliev:2008bq}. Hence, in this paper, we demonstrated that usage of the phantom field is not necessary and can be avoided by inverting the Higgs-like potential to construct a gravitating scalaron where the scalar field can possess the proper sign of the kinetic term. We explored such a scalar potential in our recent work, where the scalar field is nontrivial at the horizon, leading to the formation of a hairy black hole that bifurcated from the Schwarzschild black hole~\cite{Chew:2023olq}.

{In the presence of a non-trivial scalar field at the origin,} the gravitating scalaron also bifurcates from the Minkowski space by gaining the positive ADM mass analogous to the existence of their counterpart hairy black hole. Hence, our gravitating scalaron can be less exotic if compared to~\cite{Dzhunushaliev:2008bq}. Our gravitating scalaron can be smoothly connected to the hairy black holes in the small horizon limit. Then, we performed a systematic study of the properties of gravitating scalarons. {We found that the behavior of the profile solutions qualitatively resembles that of their counterparts in hairy black holes.} {The coefficient of the quartic term in the inverted Higgs-like scalar potential varies with positive real values and is inversely proportional to the strength of the scalar field at the origin.} In addition, our gravitating scalaron solutions are completely regular everywhere, but other classes of gravitating scalarons are not. For~instance, only the scalar field of the gravitating scalaron diverges at the origin in EsGB theory~\cite{Kleihaus:2019rbg,Kleihaus:2020qwo}, but other functions and effective stress-energy tensors are regular~everywhere. 

{Furthermore, we found that, when the massive test particle possesses a zero angular momentum, it can either move radially inward from infinity to the origin or simply remain at the origin if it is already there.} When its angular momentum is non-vanishing, it can possess a bound orbit, an~escape orbit, and also the ISCO.  However, the~gravitating scalaron does not possess a light ring, since the effective potential of the light ring does not possess an extreme point but decreases monotonically to~zero.  

Finally, we conjecture that the gravitating scalaron may inherit the instability from the hairy black holes, as they are related in the small horizon limit, implying the possible presence of the instability against the linear perturbation.

\section*{Acknowledgement}
XYC acknowledges the support provided by the starting grant from Jiangsu University of Science and Technology (JUST). We acknowledge to have a useful discussion with Hyat Huang and Jiajun Chen.

\section*{Appendix A: FJNW metric} \label{ApA}
To obtain the FJNW metric, we begin with the Einstein equation and the KG equation by setting $V(\phi)=0$,
\begin{equation}
 R_{\mu \nu} - \frac{1}{2} g_{\mu \nu} R = 2 \left(  -   \frac{1}{2} g_{\mu \nu} \nabla_\alpha \phi \nabla^\alpha \phi  + \nabla_\mu \phi \nabla_\nu \phi    \right) \,, \quad  
 \nabla_\mu \nabla^\mu \phi  = 0 \,,  \label{eom1}
\end{equation}
where we redefine the scalar field as $\phi \rightarrow \sqrt{4\pi G} \phi$. We employ the following ansatz in order to directly obtain the metric of FJNW (Equation~\eqref{FJNW}):
\begin{equation}
 ds^2 = -F_0 (r) dt^2 + \frac{1}{F_0 (r)} \left[ dr^2  + r^2 F_1(r) \left( d \theta^2+\sin^2 \theta d\varphi^2 \right)  \right] \,.
\end{equation}
Hence, the~substitution of the above metric into Equation~\eqref{eom1} yields a set of ODEs,
\begin{align}
-\frac{F''_1}{F_1}+ \frac{1}{4} \frac{F'^2_1}{F^2_1}+ \frac{F''_0}{F_0} +  \frac{F'_0 F'_1}{F_0 F_1} - \frac{5}{4} \frac{F'^2_0}{F^2_0} - 3 \frac{F'_1}{r F_1} + 2 \frac{F'_0}{r F_0}  -\frac{1}{r^2} + \frac{1}{r^2 F_1}  &=  \phi'^2 \,, \label{G11} \\ 
 \frac{1}{4} \frac{F'^2_1}{ F^2_1} - \frac{1}{4} \frac{F'^2_0}{ F^2_0} + \frac{F'_1}{r F_1} + \frac{1}{r^2} - \frac{1}{r^2 F_1} &=  \phi'^2 \,, \label{G22} \\
\frac{F''_1}{2 F_1} - \frac{1}{4} \frac{F'^2_1}{F^2_1} + \frac{1}{4} \frac{F'^2_0}{F^2_0} + \frac{F'_1}{r F_1} &= -  \phi'^2 \,, \label{G33} \\
\left(  r^2 F_1 \phi' \right)' &= 0 \,.
\end{align}
In particular, the~last ODE yields a first-order differential equation,
\begin{equation}
 \phi' = \frac{q}{r^2 F_1} \,,
\end{equation}
where $q$ is identified as the scalar~charge. 

The addition of Equations~\eqref{G22} and \eqref{G33} yields a simple ODE only for the function $F_1$,
\begin{equation}
 \frac{F''_1}{2 F_1} + 2 \frac{F'_1}{r F_1} + \frac{1}{r^2} - \frac{1}{r^2 F_1} = 0\,.
\end{equation} 
The direct integration of the above ODE yields a general solution for $F_1(r)$ using the software Maple (Maple 2023.1, Maplesoft, Waterloo), 
\begin{equation}
 F_1 (r)= 1 + \frac{C_1}{r} + \frac{C_2}{r^2} \,,
\end{equation} 
where $C_1$ and $C_2$ are the integration constants. Then, we can substitute $F_1(r)$ into \linebreak{Equation~\eqref{G33}} to obtain the following ODE for $F_0(r)$ after some algebraic simplification using Maple:
\begin{equation}
   \frac{d}{dr} \ln F_0 =  \frac{\sqrt{ -4 q^2 + C^2_2 - 4 C_1}}{r^2 + C_2 r + C_1}  \,.
\end{equation}
The above ODE can be integrated directly as follows:
\begin{equation}
    F_0 (r) =  \left( C_3 \frac{ 1 - \frac{2r+C_2}{\sqrt{C^2_2-4C_1}}}{ 1 + \frac{2r+C_2}{\sqrt{C^2_2-4C_1}}     }  \right)^{ \frac{\sqrt{ -4 q^2 + C^2_2 - 4 C_1}}{ \sqrt{C^2_2-4C_1}  } },
\end{equation}
where $C_3$ is an integration constant. We also integrate $\phi'(r)$ directly to obtain $\phi(r)$ as
\begin{equation}
 \phi (r) = q \frac{\sqrt{ -4 q^2 + C^2_2 - 4 C_1}}{ \sqrt{C^2_2-4C_1}  } \ln \left( C_4 \frac{ 1 - \frac{2r+C_2}{\sqrt{C^2_2-4C_1}}}{ 1 + \frac{2r+C_2}{\sqrt{C^2_2-4C_1}}     }  \right) \,,
\end{equation}
where $C_4$ is an integration constant. When we fix $C_2=0$ and $C_1=-b$, the~solutions become
\begin{equation}
  F_0 (r) =  \left( C_3 \frac{b-r}{r} \right)^{ \frac{\sqrt{b^2-4 q^2}}{b} }  \,, \quad
  \phi (r) = \frac{q}{b} \ln \left( C_4 \frac{b-r}{r}   \right) \,.
\end{equation}
We fix $C_3=C_4=-1$ such that $r>b$, and hence we obtain
\begin{equation}
  F_0 (r) =  \left( 1-\frac{b}{r} \right)^{ \frac{\sqrt{b^2-4 q^2}}{b} }  \,, \quad
  \phi (r) = \frac{q}{b} \ln \left( 1-\frac{b}{r}   \right) \,.
\end{equation}
Note that the solutions diverge when $r=b$. In~particular, the scalar field diverges as $\ln r$ when $r \rightarrow 0$. Then, one could expand the function $F_0(r)$ in the limit $r \rightarrow \infty$,
\begin{equation}
 \left( 1-\frac{b}{r} \right)^{ \frac{\sqrt{b^2-4 q^2}}{b} } \approx 1 - \frac{\sqrt{b^2-4 q^2}}{r} + O\left( \frac{1}{r^2} \right)\,,
\end{equation}
and then one could read off the ADM mass directly as
\begin{equation}
 M = \frac{1}{2} \sqrt{b^2-4 q^2} \,. 
\end{equation} 
One can rewrite the solutions as
\begin{equation}
  F_0 (r) =  \left( 1-\frac{b}{r} \right)^\gamma  \,, \quad
  \phi (r) = \frac{q}{b} \ln \left( 1-\frac{b}{r}   \right) \,,
\end{equation}
where we have introduced a constant $\gamma=2M/b$. Finally, we have verified that the solutions can satisfy Equations~\eqref{G11}--\eqref{G33}.

\section{Appendix B: Hairy Black Hole} \label{ApB}

We briefly revisited the properties of the hairy black hole that was constructed recently in our paper~\cite{Chew:2023olq}. The~solutions to the hairy black hole could be obtained numerically by solving Equations~\eqref{EFE} and \eqref{KGeqn} with COLSYS and bvp4c. In~the numerics, the~solutions are required to be regular at the horizon $r_H$, which can be represented by the following series expansions:
\begin{align}
 m(r) &= \frac{r_H}{2}+ m_1 (r-r_H) + O\left( (r-r_H)^2 \right) \,, \label{m_ex} \\
\sigma(r) &= \sigma_H + \sigma_1   (r-r_H) + O\left( (r-r_H)^2 \right)  \,, \\
 \phi(r) &= \phi_H +  \phi_{H,1}  (r-r_H) + O\left( (r-r_H)^2 \right) \label{p_ex} \,,
\end{align} 
where
\begin{equation}
   m_1 = 4 \pi G r^2_H  V(\phi_H)  \,, \quad  \sigma_1 = -  4 \pi G r_H \phi^2_{H,1} \,, \quad   \phi_{H,1}= \frac{r_H \frac{d V(\phi_H)}{d \phi}}{1-8 \pi G r_H^2 V(\phi_H)}  \,.
\end{equation}  
Here, $\sigma_H$ and $\phi_H$ denote the values of $\sigma$ and $\phi$ at the horizon, respectively. It should be emphasized that the denominator of $\phi_{H,1}$ must fulfill the condition $1-8 \pi G r_H^2 V(\phi_H) \neq 1$ in order to maintain $\sigma(r)$ and $\phi(r)$ as finite at the horizon. Meanwhile, the~hairy black hole shares the same asymptotic behavior with the gravitating scalaron, which is given by Equations~\eqref{in1}--\eqref{in3}.

{We can analyze the characteristics of the horizon for the hairy black holes using two fundamental quantities: the horizon area $A_H$ and the Hawking temperature $T_H$:}
\begin{equation}
A_H = 4 \pi r^2_H \,, \quad T_H = \frac{1}{4 \pi} N'(r_H) e^{-\sigma_H}   \,.
\end{equation}
The Ricci scalar $R$ and Kretschmann scalar $K=R_{\alpha \beta \gamma \delta} R^{\alpha \beta \gamma \delta}$ of the hairy black holes are given as follows:
\begin{align}
 R &= -N'' + \frac{3 r \sigma'-4}{r} N' + \frac{2 \left( 2 r N \sigma' - N +1 + r^2 N\sigma'' - r^2 N \sigma'^2  \right)}{r^2} \,, \\
 K &= \left(  3 \sigma' N' + 2 N \sigma'' - N'' -2 N \sigma'^2  \right)^2    +  \frac{2}{r^2} \left( N'-2 N \sigma'  \right)^2 + \frac{2 N'^2}{r^2} + \frac{4 (N-1)^2}{r^4}  \,.
\end{align}
At the horizon, the~finite functions of $R$ and $K$ with a few leading orders using series expansion are:
\begin{align}
  R &=  -\frac{2 m_1 \left( 3 r_H \sigma_1 -2 \right)}{r^2_H} + \frac{3 \sigma_1 + 4 m_2}{r_H} +  O\left( (r-r_H) \right)  \,, \label{R_ex}  \\
  K &=  \frac{16 m^2_2}{r^2_H} - \frac{8 \left(  -2 + 6 m_1 \sigma_1 r_H + 4 m_1 - 3 \sigma_1 r_H  \right)}{r^3_H}   \nonumber \\
& \quad + \frac{1}{r^4_H} \left(  12-32 m_1 + 48 m^2_1 \sigma_1 r_H + 36 m^2_1 \sigma^2_1 r^2_H + 32 m^2_1 + 9 \sigma^2_1 r^2_H \right. \nonumber \\   
& \qquad \qquad \qquad \left.  + 12 \sigma_1 r_H - 36 \sigma^2_1 r^2_H m_1 - 48 m_1 \sigma_1 r_H  \right) +  O\left( (r-r_H) \right)  \,, \label{K_ex}
\end{align}
{where $m_2$ denotes the coefficient of the second-order term in Equation \eqref{m_ex}. While the Ricci scalar of the Schwarzschild black hole is zero, its Kretschmann scalar is expressed by $K=12r_H^2/r^6$.}

We present several numerical results. {A branch of hairy black holes bifurcates from the Schwarzschild black hole when $\phi_H$ is nontrivial at the horizon.} Figure~\ref{plot_M_TH}a shows that the ADM mass of the hairy black hole increases with the increase in $\phi_H$. Figure~\ref{plot_M_TH}b also shows a similar feature for the Hawking temperature of the hairy black hole. In~principle, $\phi_H$ can take any positive real values. Figure~\ref{plot_Ttt} shows the violation of the weak energy condition, which is described by the scaled energy density $(8\pi G/\mu) \rho=- (8\pi G/\mu) T^t\,_t$.

\begin{figure}
\centering 
\mbox{
(\textbf{a})
\includegraphics[angle = 0,scale=0.55]{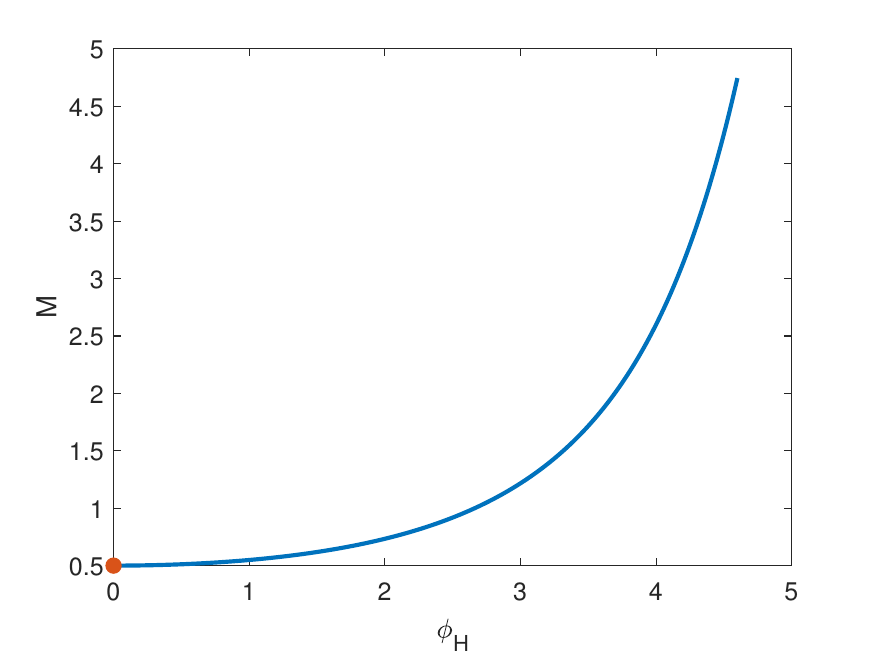}
(\textbf{b})
\includegraphics[angle = 0,scale=0.55]{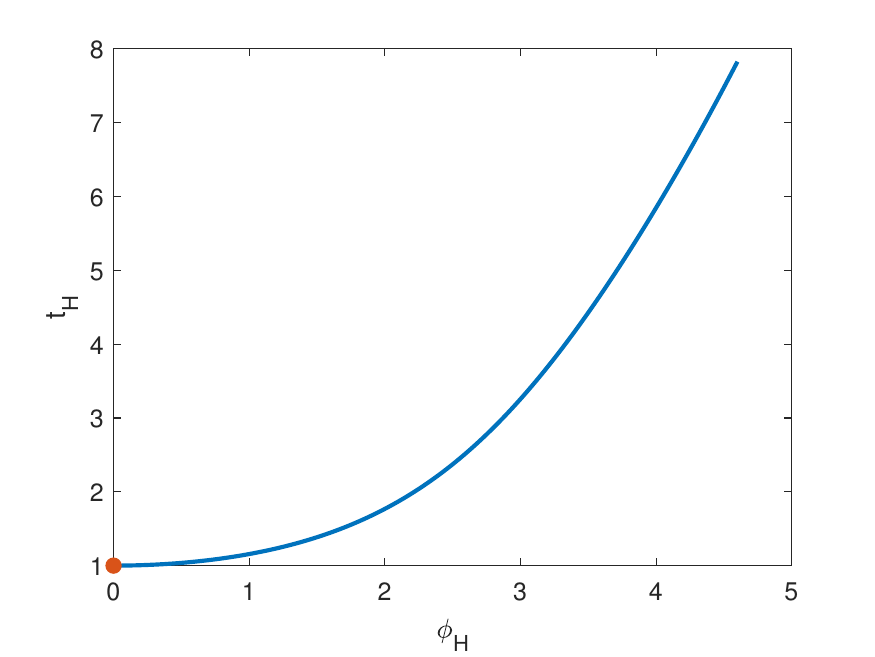}
}
\caption{The characteristics of the hairy black hole, with~$r_H=1$, are illustrated by: (\textbf{a}) the ADM mass versus $\phi_H$; and (\textbf{b}) the Hawking temperature versus $\phi_H$. The~orange dot signifies the value of the Schwarzschild black hole.}
\label{plot_M_TH}
\end{figure}
\vspace{-12pt}

\begin{figure}
\mbox{
\includegraphics[angle = 0,scale=0.55]{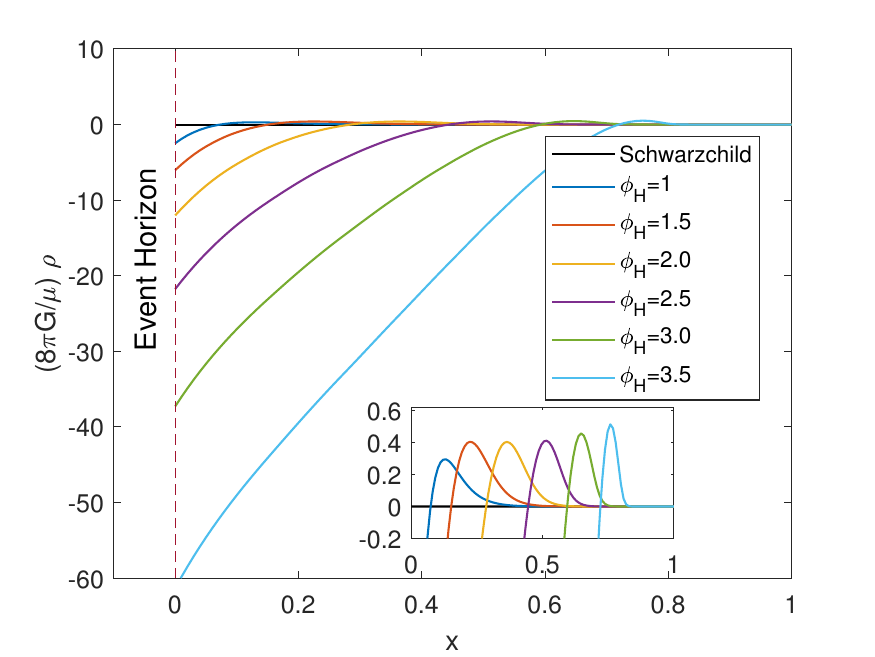}
}
\caption{The violation of the weak energy condition of the hairy black hole in the compactified coordinate $x=1-r_H/r$ for various values of $\phi_H$ with $r_H=1$.}
\label{plot_Ttt}
\end{figure}

\end{document}